\documentclass[12pt]{article}
\usepackage{latexsym,amsmath,amssymb, graphicx, subfigure}
\renewcommand{\thefootnote}{\fnsymbol{footnote}}

\usepackage{mathrsfs }
\usepackage{amsmath}
\usepackage{amsfonts}

\numberwithin{equation}{section}
	
\def\doubleset#1#2{\bgroup%
\def\doit#1#2{%
\setbox\dblsetbox=\hbox{$\cstyle #1$}%
\raise#2\ht\dblsetbox\copy\dblsetbox%
\hskip-\wd\dblsetbox%
\raise-#2\ht\dblsetbox\box\dblsetbox}%
\mathchoice%
{\def\cstyle{\displaystyle}\doit#1#2}%
{\def\cstyle{\textstyle}\doit#1#2}%
{\def\cstyle{\scriptstyle}\doit#1#2}%
{\def\cstyle{\scriptscriptstyle}\doit#1#2}\egroup}
\def\underarrow#1{\vbox{\ialign{##\crcr$\hfil\displaystyle
 {#1}\hfil$\crcr\noalign{\kern1pt\nointerlineskip}$\longrightarrow$\crcr}}}

\def\IL{\relax{\rm I\kern-.18em L}}
\def\IH{\relax{\rm I\kern-.18em H}}
\def\IR{{\mathbb R}}
\def\IB{\relax{\rm I\kern-.18em B}}
\def\ID{\relax{\rm I\kern-.18em D}}
\def\IE{\relax{\rm I\kern-.18em E}}
\def\IF{\relax{\rm I\kern-.18em F}}
\def\IZ{{\mathbb Z}}
\def\IG{\relax\hbox{$\inbar\kern-.3em{\rm G}$}}
\def\IGa{\relax\hbox{${\rm I}\kern-.18em\Gamma$}}
\def\IH{\relax{\rm I\kern-.18em H}}
\def\II{\relax{\rm I\kern-.18em I}}
\def\IK{\relax{\rm I\kern-.18em K}}
\def\IP{\relax{\rm I\kern-.18em P}}
\def\IQ{\relax\hbox{$\inbar\kern-.3em{\rm Q}$}}

\def\hat{\widehat}
\def\CM {{\cal M}}
\def\CN {{\cal N}}

\def\CD {{\cal D}}

\def\CQ{{\cal Q}}

\def\inbar{\,\vrule height1.5ex width.4pt depth0pt}

\def\Xminus{{X \backslash D}}
\def\spin{{{\rm Spin}^c}}
\def\lama{{\lambda'}}

\newbox\dblsetbox

\newcommand{\dirac}{D\mspace{-13mu}/\mspace{4mu}}

\newlength{\extraspace}
\newlength{\extraspaces}
\newcommand{\be}{\begin{equation}
\addtolength{\abovedisplayskip}{\extraspaces}
\addtolength{\belowdisplayskip}{\extraspaces}
\addtolength{\abovedisplayshortskip}{\extraspace}
\addtolength{\belowdisplayshortskip}{\extraspace}}
\newcommand{\ee}{\end{equation}}

\newcommand{\ba}{\begin{eqnarray}
\addtolength{\abovedisplayskip}{\extraspaces}
\addtolength{\belowdisplayskip}{\extraspaces}
\addtolength{\abovedisplayshortskip}{\extraspace}
\addtolength{\belowdisplayshortskip}{\extraspace}}
\newcommand{\ea}{\end{eqnarray}}

\newcommand{\bd}{\begin{displaymath}
\addtolength{\abovedisplayskip}{\extraspaces}
\addtolength{\belowdisplayskip}{\extraspaces}
\addtolength{\abovedisplayshortskip}{\extraspace}
\addtolength{\belowdisplayshortskip}{\extraspace}}
\newcommand{\ed}{\end{displaymath}}

\newcounter{saveeqn}

\newcommand{\newsection}[1]{
\vspace{12mm} \pagebreak[3] \addtocounter{section}{1}
\setcounter{equation}{0} \setcounter{subsection}{0}
\noindent{\bf \thesection. #1} \nopagebreak
\medskip
\nopagebreak
\addcontentsline{toc}{section}{\thesection. #1}}

\newcommand{\newsubsection}[1]{
\vspace{0.8cm} \pagebreak[3] \addtocounter{subsection}{1}
\setcounter{subsubsection}{0}
\noindent{ \it \thesubsection. #1} \nopagebreak \vspace{2mm}
\nopagebreak
\addcontentsline{toc}{subsection}{\thesubsection. #1}}

\begin{document}
\addtolength{\baselineskip}{1.5mm}

\thispagestyle{empty}

\vbox{} \vspace{1.3cm}

\begin{center}
\centerline{\LARGE{Supersymmetric Surface Operators,}}
\medskip
\centerline{\LARGE{Four-Manifold Theory and Invariants}}      
\medskip       
\centerline{\LARGE{in Various Dimensions}}      
\medskip

\vspace{0.9cm}

{\bf{Meng-Chwan~Tan} \footnote{email: tan@ias.edu}}
\\[1mm]
{\it Department of Physics,
National University of Singapore \\
Singapore 119260}\\[0 mm]
\end{center}

\vspace{0.9cm}

\centerline{\bf Abstract}\smallskip \noindent

We continue our program initiated in~\cite{u-plane SO} to consider supersymmetric surface operators in a topologically-twisted $\CN =2$ pure $SU(2)$ gauge theory, and apply them to the study of four-manifolds and related invariants. Elegant physical proofs of various seminal theorems in four-manifold theory obtained by Ozsv\'ath-Szab\'o~\cite{OS,OS 2} and Taubes~\cite{CT 1}, will be furnished. In particular, we will show that Taubes' groundbreaking and difficult result -- that the ordinary Seiberg-Witten invariants are in fact the Gromov invariants which count pseudo-holomorphic curves embedded in a symplectic four-manifold $X$ -- nonetheless lends itself to a simple and concrete physical derivation in the presence of ``ordinary'' surface operators.  As an offshoot, we will be led to several interesting and mathematically novel identities among the Gromov and  ``ramified'' Seiberg-Witten invariants of $X$, which in certain cases, also involve the instanton and monopole Floer homologies  of its three-submanifold. Via these identities, and a physical formulation of the ``ramified'' Donaldson invariants of four-manifolds with boundaries, we will uncover completely new and economical ways of deriving and understanding various important mathematical results concerning (i) knot homology groups from ``ramified'' instantons  by Kronheimer-Mrowka~\cite{dym}; and (ii) monopole Floer homology and Seiberg-Witten theory on symplectic four-manifolds by Kutluhan-Taubes~\cite{CT 1,SWF}. Supersymmetry, as well as other physical concepts such as $R$-invariance, electric-magnetic duality, spontaneous gauge symmetry-breaking and localization onto supersymmetric configurations in topologically-twisted quantum field theories, play a pivotal role in our story.

\newpage

\renewcommand{\thefootnote}{\arabic{footnote}}
\setcounter{footnote}{0}

\tableofcontents 

\newsection{Introduction And Summary}

Supersymmetric surface operators in a topologically-twisted $\CN =2$ pure $SO(3)$ or $SU(2)$ gauge theory have recently been analyzed in detail in~\cite{u-plane SO}, where, among other things, concrete physical proofs of various seminal theorems in four-dimensional geometric topology obtained by Kronheimer and Mrowka in~\cite{structure, KM1, KM2}, were also furnished.  For example, it was shown in~\cite{u-plane SO} that the Kronheimer-Mrowka result of~\cite{structure} -- which identifies the ``ramified'' Donaldson invariants as the ordinary Donaldson invariants of  an ``admissible'' four-manifold $X$ with $b^+_2 > 1$ -- is a direct consequence of a required modular invariance over the $u$-plane in the presence of nontrivially-embedded surface operators. It was also shown in~\cite{u-plane SO} that a generalization of the Thom conjecture  proved by Kronheimer and Mrowka in~\cite{structure}  --  which leads to a minimal genus formula for embedded surfaces of non-negative self-intersection in $X$ -- is a direct result of the $R$-invariance of the correlation functions in the microscopic non-abelian gauge theory which correspond to the (``ramified'') Donaldson invariants of $X$.

In this paper, we continue the program initiated in~\cite{u-plane SO}; we consider arbitrarily-embedded surface operators in a topologically-twisted $\CN =2$ pure $SU(2)$ gauge theory, and apply them to the study of four-manifolds and invariants in two, three and four dimensions. The plan and results of our work can be summarized as follows. 


In $\S$2, we will review various aspects of the topologically-twisted  $\CN =2$ pure $SU(2)$ gauge theory on $X$ with arbitrarily-embedded surface operators, and the corresponding physical interpretations of the ``ramified'' Donaldson and Seiberg-Witten invariants and their associated moduli spaces, all of which will be useful and relevant to our arguments and computations in the later sections. 

In $\S$3, with the aid of key results computed in~\cite{u-plane SO}, we will  furnish physical proofs of various seminal theorems in four-dimensional geometric topology obtained by Ozsv\'ath-Szab\'o in~\cite{OS,OS 2}; in particular, we will physically demonstrate a minimal genus formula obtained earlier in~\cite{OS} for embedded surfaces of \emph{negative} self-intersection. $R$-invariance and electric-magnetic duality underlie our proofs in this section.  

In $\S$4, we will present an elegant physical derivation of Taubes' stunning result in~\cite{CT 1}, which identifies the Seiberg-Witten invariants as the Gromov invariants on a symplectic four-manifold with $b^+_2 > 1$. The crucial ingredients in this derivation are supersymmetry, $R$-invariance, electric-magnetic duality, spontaneous gauge symmetry-breaking and localization onto supersymmetric configurations in topologically-twisted quantum field theories. In essence, one can understand Taubes' result to be a consequence of the scale invariance of a particular instanton sector of the topologically-twisted gauge theory in the presence of  ``ordinary'' curved surface operators which wrap pseudo-holomorphic curves embedded in the symplectic four-manifold.  

In $\S$5, we will explore the mathematical implications of the underlying physics. We will compute --  using certain intermediate results obtained in $\S$3 and $\S$4 -- various mathematically novel identities involving the Gromov and (``ramified'') Seiberg-Witten invariants of a symplectic four-manifold with $b^+_2 > 1$. These identities, which one can understand to exist because of $R$-invariance, are also found to be consistent with more general theorems established in the mathematical literature. In addition, for symplectic $X = M \times {{\bf S}^1}$, where $M$ is a closed, oriented three-submanifold, we will show -- via a supersymmetric quantum mechanical interpretation of the topological gauge theory with surface operators -- that a knot homology conjecture proposed by Kronheimer and Mrowka in~\cite{dym} ought to hold on purely physical grounds, and that the Gromov invariant of $X$ is given by the Euler characteristic of the instanton Floer homology of $M$. In turn, because the Euler characteristic of the instanton Floer homology of $M$ is given by the Casson-Walker-Lescop invariant of $M$, the Gromov invariant of $X$ is zero if $b^+_2(X) > 3$. We will also derive, amidst other things,  an interesting relation between the instanton and monopole Floer homologies of $M$, and a novel identity between the Seiberg-Witten invariants of $M$. Last but not least, we will formulate ``ramified'' generalizations of various formulas presented by Donaldson and Atiyah in~\cite{Donald, Weyl} relating ordinary Donaldson and Floer theory on four-manifolds with boundaries, in terms of ``three-one branes''.

In $\S$6, we will generalize our computations in $\S$4 to involve multiple surface operators which are \emph{disjoint}. This will allow us to physically derive Taubes' result in all generality.

In $\S$7, the final section, by further applying our physical insights and results obtained hitherto, we will first provide --  via the topological gauge theory with nontrivially-embedded surface operators on a general four-manifold with boundaries -- a physical derivation of certain key properties of knot homology groups from ``ramified'' instantons defined and proved by Kronheimer and Mrowka in~\cite{dym}. Then, via the identities obtained in $\S$5, and certain key relations computed in $\S$3 and $\S$4, we will re-derive various important mathematical results concerning the monopole Floer homology of three-manifolds and Seiberg-Witten theory on symplectic four-manifolds.

\medskip
This paper is dedicated to See-Hong, whose strength, courage and optimism in the face of grave adversity have made this otherwise impossible endeavor, possible.

\newpage
\newsection{Surface Operators And The ``Ramified'' Donaldson And Seiberg-Witten Invariants}

In this section, we will present some background material that will be necessary for a coherent, self-contained understanding of the main discussions in this paper. We will be brief in our exposition, although concepts deemed to play a crucial role will be reviewed in greater detail.

 \newsubsection{Embedded Surfaces And The ``Ramified'' Donaldson Invariants}

Let us first review the mathematical definition of embedded surfaces and the ``ramified'' Donaldson invariants by Kronheimer and Mrowka (henceforth denoted as KM) in~\cite{structure}.  To this end, let $X$ be a smooth, compact, simply-connected, oriented 
four-manifold with Riemannian metric $\bar g$, and let $E\rightarrow X$
be an $SO(3)$-bundle over $X$ (i.e., a rank-three real vector
bundle with a metric).

\bigskip\noindent{\it Embedded Surfaces}

An embedded surface $D$ is characterized by a two-submanifold of $X$  that is a complex curve of genus $g$ and self-intersection number $D^2$. Consider  the case where the second Stiefel-Whitney class $w_2(E) =0$; the structure group of $E$ can then be lifted to its $SU(2)$ double-cover. In the neighborhood of $D$, one can choose a decomposition of  $E$ as
\be
E = {L} \oplus {L}^{-1},
\label{decompose}
\ee
where $L$ is a complex line bundle over $X$. In the presence of $D$, the connection matrix of $E$ restricted to $X \backslash  D$ (in the \emph{real} Lie algebra) will look  like 
\be
A = \alpha  d \theta + \cdots,
\label{connection}
\ee
where $\alpha$ is a real number valued in the generator 
\be
\begin{bmatrix} 
 1 & \ 0\\ 
0 & -1
\end{bmatrix}
\label{frak t}
\ee
of the Cartan subalgebra $\frak t$, $\theta$ is the angular variable of the coordinate $z = re^{i \theta}$ of the plane normal to $D$, and the ellipses refer to the ordinary terms that are regular near $D$. Notice that since $d\theta = i dz / z$, the connection is singular as $z \to 0$, i.e.,  as one approaches $D$. In any case, the singularity in the connection induces the following gauge-invariant holonomy 
\be
\textrm{exp} ( 2 \pi  \alpha) 
\label{holonomy}
\ee
around any small circle that links $D$. Hence, if the holonomy is trivial, we are back to considering ordinary connections on $E$. Therefore, $\alpha$ effectively takes values in  $\mathbb T$, the maximal torus of the gauge group with Lie algebra $\frak t$. As we shall see shortly, this mathematical definition of embedded surfaces will coincide with our (more general) physical definition of surface operators. 

\bigskip\noindent{\it The ``Ramified'' Donaldson Invariants}

In analogy with the original formulation of Donaldson theory~\cite{donaldson},  KM introduced the notion of ``ramified'' Donaldson invariants -- i.e., Donaldson invariants of $X$ with an embedded surface $D$. According to KM~\cite{structure,KM2}, the ``ramified'' Donaldson polynomials $\CD'_E$ can be defined as polynomials
on the homology of $\Xminus$ with real coefficients:\footnote{To be precise, KM actually defines the ``ramified'' Donaldson polynomials to be the map $\CD'_E : \textrm{Sym}[H_0(\Xminus,\IR) \oplus H_2(\Xminus,\IR)] \otimes \wedge*H_1(\Xminus, \IR)  \rightarrow \IR$. However, their definition can be truncated as shown, in accordance with Donaldson's original formulation in~\cite{donaldson}.}
\be
\CD'_E: H_0(\Xminus,\IR)
\oplus H_2(\Xminus,\IR) \rightarrow \IR .
\ee

Assigning degree  $4$ to
$p\in H_0(\Xminus,\IR)$ and $2$ to $S \in H_2(\Xminus,\IR)$,
the degree $s$  polynomial  may be expanded as
\be{
\CD'_E(p,S) = \sum_{2m + 4 t = s} S^m p^t d^{k'}_{m,t},
\label{Donaldson polynomial}
}\ee
where $s$ is the dimension of the moduli space ${\cal M}'$
of gauge-inequivalent classes of anti-self-dual  connections  on $E$ restricted to $X \backslash D$ with first Pontrjagin number $p_1(E)$ and instanton number $k' = - \int_{X \backslash D} p_1(E) /4$. The numbers $d^{k'}_{m,t}$ -- in other words, the ``ramified'' Donaldson invariants of $X$ -- can be defined as in  Donaldson theory in terms of intersection theory on the moduli space ${\cal M}'$; for maps 
\begin{eqnarray}
\label{map of defn}
p \in  H_0(\Xminus,\IR) & \rightarrow & \Omega^0(p) \in H^4(\CM'), \nonumber \\
S \in H_2(\Xminus,\IR)  & \rightarrow & \Omega^2 (S) \in H^2 (\CM'), 
\end{eqnarray}
the ``ramified'' Donaldson invariants can be written as
\be
d^{k'}_{m,t} = \int_{\CM'} [\Omega^0(p)]^t \wedge \Omega^2 (S_{i_1}) \wedge \dots \wedge \Omega^2(S_{i_m}).
\label{ram DW}
\ee

Moreover, one can also
package the ``ramified'' Donaldson polynomials into
a generating function: by summing over all topological
types of bundle $E$ with fixed $\xi=w_2(E)$ (where $\xi$ may be non-vanishing in general) but varying $k'$, the generating function can be defined as 
\be
{\bf Z}'_{\xi, \bar g}(p,S) =
\sum_{k'}\sum_{m \geq 0, t\geq 0} {S^m \over  m!}{ p^t \over  t!} d^{k'}_{m,t}.
\label{Donaldson generating function}
\ee
Clearly, ${\bf Z}'_{\xi, \bar g}$ depends on the class $w_2(E)$
 but not on the instanton number $k'$  (as this
has been summed over).  In analogy with the ordinary case, one can also define the ``ramified'' Donaldson series as
\be
\label{series}
{\mathscr D}' _\xi (S) = \left(1 + {1\over 2} {\partial \over \partial p}\right) \cdot {\bf Z}'_{\xi, \bar g}(p,S) \vert_{p=0}. 
\ee

\bigskip\noindent{\it About The Moduli Space Of ``Ramified'' Instantons}

Another relevant result by KM is the following. Assuming that there are no reducible connections on $E$ restricted to $X \backslash D$,  ${\cal M}'$ -- which we will hereon refer to as the moduli space of ``ramified'' instantons -- will be a smooth manifold  of finite dimension
\be
s = 8k - {3 \over 2}(\chi + \sigma) + 4 {l} - 2( g-1)
\ee
for $\it any$ nontrivial value of $\alpha$.  Here, $\chi$ and $\sigma$ are the Euler characteristic and signature of $X$, and for $\xi =0$, the integer $k$ is given by 
\be
k = -{1 \over 8 \pi^2} \int_X \textrm{Tr} \hspace{0.1cm} F \wedge F,
\ee  
where $F$ is the curvature of the bundle $E$ over $X$, and $\textrm{Tr}$ is the trace in the two-dimensional representation of $SU(2)$.  The integer $l$ -- called the monopole number by KM -- is given by
\be
{l} = - \int_D c_1(L).
\ee
Here, $c_1(L) = - F_L / 2\pi$, where $F_L$ is the curvature of $L$; thus, $l$ measures the degree of the reduction of $E$ near $D$. 

As shown in~\cite{u-plane SO}, $l$ will depend explicitly on $\alpha$ because the singular term in the connection $A$ will result in a singularity proportional to $\alpha$ along $D$ in the field strength (extended over $D$).  Likewise, $k$ will also depend explicitly on $\alpha$. Thus, the invariance of $s$ must mean that both $l$ and $k$ will vary with $\alpha$  in such a way as to keep it fixed for any nontrivial value of $\alpha$.  

\bigskip\noindent{\it Topological Invariance Of ${\bf Z}'_{\xi, \bar g}$}

Let $b^+_2$ denote the self-dual part of the second Betti number of $X$. According to KM (see $\S$7 of~\cite{KM2}), if $b^+_2  > 1$, ${\bf Z}'_{\xi, \bar g}$ is independent of the metric $\bar g$ and hence,  just like the generating function of the ordinary Donaldson invariants, defines invariants of the smooth structure of $X$. This is consistent with the fact that for $b^+_2 \geq 3$ (and $b_1 =0$), the ``ramified'' Donaldson invariants can be expressed solely in terms of the ordinary Donaldson invariants (see Theorem 5.10 of~\cite{structure}, and its physical proof in $\S$8 of~\cite{u-plane SO}).  

However, if $b_2^+=1$, we run into the phenomenon of chambers; ${\bf Z}'_{\xi, \bar g}$ will jump as we move across a ``wall'' in the space of metrics on $X$. (This phenomenon was demonstrated via a purely physical approach in $\S$6 of~\cite{u-plane SO}.)

\newsubsection{Surface Operators In Pure $SU(2)$ Theory With ${\CN} =2$ Supersymmetry}

\medskip\noindent{\it Supersymmetric Surface Operators}

We would like to define surface operators along $D$ which are compatible with ${\cal N} =2$ supersymmetry.  In other words,  they should be characterized by solutions to the supersymmetric field configurations of the underlying gauge theory on $X$ that are singular along $D$. 

In order to ascertain what these solutions are, first note that any supersymmetric field configuration of a theory must obey the conditions implied by setting the supersymmetric variations of the fermions to zero. In the original (untwisted) theory without surface operators, this implies that any supersymmetric field configuration must obey $F =0$ and $\nabla_\mu a = 0$, where $a$ is a scalar field in the ${\cal N} = 2$ vector multiplet~\cite{Marcos}. Let us assume for simplicity the trivial solution $a=0$ to the condition $\nabla_\mu a =0$ (so that the relevant moduli space is non-singular); this means that any supersymmetric field configuration must be consistent with $\it{irreducible}$ flat connections on $X$ that obey $F=0$. Consequently, any surface operator along $D$ that is supposed to be supersymmetric and compatible with the underlying ${\cal N} =2$ supersymmetry,  ought to correspond to a $\it{flat}$ irreducible connection on $E$ restricted to $X \backslash D$ which has the required singularity  along $D$.\footnote{This prescription of considering connections on the bundle $E$ restricted to $X \backslash D$ whenever one inserts a surface operator that introduces a field singularity along $D$,  is just a two-dimensional analog of the prescription one adopts when inserting an 't Hooft loop operator in the theory. See $\S$10.1 of~\cite{QFT2} for a detailed explanation of the latter.}  Let us for convenience choose  the singularity of the connection along an oriented $D$ to be of the form shown in (\ref{connection}). Then,  since $d (\alpha d \theta) = 2 \pi \alpha \delta_D$, where $\delta_D$ is a delta two-form Poincar\'e dual to $D$  with support in a tubular neighborhood of $D$~\cite{bott-tu},  our surface operator will equivalently correspond to a $\it{flat}$ irreducible connection on a bundle $E'$ over $X$ whose field strength is $F' = F - 2\pi \alpha \delta_D$, where $F$ is the field strength of the bundle $E$ over $X$.\footnote{To justify this statement, note that the instanton number $\tilde k$ of the bundle $E$ over $X \backslash D$ is (in the mathematical convention)  given by $\tilde k = k +  2 \alpha {l} - \alpha^2 D \cap D$, where $k$ is the instanton number of the bundle $E$ over $X$ with curvature $F$, and $l$ is the  monopole number  (cf.~eqn.~(1.7) of~\cite{KM1}). On the other hand,  the instanton number $k'$ of the bundle $E'$ over $X$ with curvature $F' = F - 2\pi \alpha \delta_D$ is (in the physical convention) given by $k' = - {1\over 8 \pi^2} \int_X \textrm {Tr} F' \wedge F' = k +  2 \alpha {l} - \alpha^2 D \cap D$. Hence, we find that the expressions for $\tilde k$ and $k'$ coincide, reinforcing the notion that the bundle $E$ over $X \backslash D$ can be equivalently interpreted as the bundle $E'$ over $X$. Of course, for $F'$ to qualify as a nontrivial field strength, $D$ must be a homology cycle of $X$, so that $\delta_D$ (like $F$) is in an appropriate cohomology class  of $X$.}  In other words, a supersymmetric surface operator will correspond to a gauge field solution over  $X$ that satisfies 
\be
F = 2 \pi \alpha \delta_D
\label{surface operator}
\ee
along $D$.  Indeed,  the singular term in $A$ of (\ref{connection}) that is associated with the inclusion of an embedded surface, is such a solution. Thus, our physical definition of supersymmetric surface operators coincides with the mathematical definition of embedded surfaces. 

Some comments on (\ref{surface operator}) are in order. Note that even though $\alpha$ is formally defined in (\ref{connection}) to be $\frak t$-valued, we saw that it effectively takes values in the maximal torus $\mathbb T$. Since $\mathbb T = {\frak t} / \Lambda_{\textrm{cochar}}$, where $\Lambda_{\textrm{cochar}}$ is the cocharacter lattice of the underlying gauge group,  (\ref{surface operator}) then appears to be unnatural as one is free to subject $F$ to a shift by an element of $\Lambda_{\textrm{cochar}}$. This can be remedied by lifting $\alpha$ in (\ref{surface operator}) from ${\frak t} / \Lambda_{\textrm{cochar}}$ to ${\frak t}$. Equivalently, this corresponds to a choice of an extension of the bundle $E$  over $D$ -- something that was implicit in our preceding discussion. 

\bigskip\noindent{The ``Quantum'' Parameter $\eta$}

With an extension of the bundle $E$ over $D$, the restriction of the field strength $F$ to $D$ will be $\frak t$-valued. Hence, we roughly have an abelian gauge theory in two dimensions along $D$. As such, one can generalize the physical definition of the surface operator, and introduce a two-dimensional theta-like angle $\eta$ as an additional ``quantum'' parameter which enters in the Euclidean path-integral via the phase
\be
\textrm{exp}\left( 2 \pi i \hspace{0.00cm} \textrm{Tr} \hspace{0.1cm}  \eta \frak m \right),
\label{eta term}
\ee
where $\frak m = \int_D F/ 2 \pi$. Since $F$ restricted to $D$ is $\frak t$-valued, and since the monopole number $l =  \int_D F_L / 2\pi$ is an integer, it will mean that $\frak m$ must take values in the subset of diagonal, traceless $2 \times 2$ matrices -- which generate the maximal torus $\mathbb T$ -- that have \emph{integer} entries only; i.e., ${\frak m} \in \Lambda_{\textrm{cochar}}$. Also, values of $\eta$  that correspond to a nontrivial phase must be such that $\textrm{Tr} \hspace{0.1cm} \eta \frak m$ is \emph{non-integral}. Because $\textrm{Tr} \hspace{0.1cm}  {\frak m}' {\frak m}$ is an integer if ${\frak m}' \in \Lambda_{\textrm{cochar}}$, it will mean that  $\eta$  must takes values in ${\frak t} / \Lambda_{\textrm{cochar}} = \mathbb T$. Just like $\alpha$, one can shift $\eta$ by an element of $\Lambda_{\textrm{cochar}}$ whilst leaving the theory invariant.\footnote{This characteristic of $\eta$ is consistent with an $S$-duality in the corresponding, low-energy  effective  abelian theory which maps  $(\alpha, \eta) \to (\eta, - \alpha)$~\cite{mine}.}  Note that modular invariance requires that $\eta$ be set to \emph{zero} if the surface operator is\emph{ nontrivially-embedded}; this condition is a crucial ingredient in the physical proof of KM's relation between the ``ramified'' and ordinary Donaldson invariants in~\cite{u-plane SO}.

\bigskip\noindent{\it A Point On Nontrivially-Embedded Surface Operators}

  More can also be said about the ``classical'' parameter $\alpha$ as follows. In the case when the surface operator is trivially-embedded in $X$ -- i.e., $X= D' \times D$ and the normal bundle to $D$ is hence trivial -- the self-intersection number
\be
D \cap D = \int_{X} \delta_D \wedge \delta_D
\label{DcapD}
\ee
vanishes. On the other hand, for a nontrivially-embedded surface operator supported on $D \subset X$, the normal bundle is nontrivial, and the intersection number is non-zero. The surface operator is then defined by the gauge field with singularity in (\ref{connection}) in each normal plane.

When the surface operators are nontrivially-embedded, there is a condition on the allowed gauge transformations that one can invoke in the physical theory~\cite{Gukov-Witten}. Let us explain this for when the underlying gauge group is $U(1)$ with gauge bundle $L$. Since there is a singularity of $2 \pi \alpha \delta_D$ in the abelian field strength $F_L$ restricted to $D$, we find, using (\ref{DcapD}), that $\int_D F_L/2 \pi  = \alpha  D \cap D \ \textrm{mod}  \ \mathbb Z$. Since $\int_D F_L/2 \pi = l$ is always an integer, we must have
\be
\alpha  D \cap D \in \mathbb Z.
\label{intersection number}
\ee

In fact, underlying the integrality of $l$ is actually  the condition $c_1(L) \in H^2(X, \IZ)$. This implies that for \emph{any} integral homology 2-cycle $U \subset X$ (assuming, for simplicity, that $H_2(X, \mathbb Z)$ is torsion-free), $-c_1(L) [U] = \int_U F_L / 2\pi = \alpha (U \cap D) \ {\rm mod} \ \IZ$ is always an integer; in other words, we must have
\be
\alpha (U \cap D) \in \mathbb Z.
\label{condition}
\ee

Now consider a gauge transformation -- in the normal plane -- by the following $U(1)$-valued function 
\be
(r, \theta) \to \textrm{exp} (\theta u),
\label{twisted gauge tx}
\ee
where $u \in {\frak u}(1)$;  its effect is to shift $\alpha \to \alpha + u$ whilst leaving the holonomy ${\rm exp}(2 \pi \alpha)$ of the abelian gauge connection around a small circle linking $D$ which underlies the \emph{effective} ``ramification'' of the theory, \emph{unchanged}. Clearly, the only gauge transformations of this kind that can be globally-defined along $D$, are those whereby the corresponding shifts in $\alpha$ are compatible with  (\ref{condition}). For effectively nontrivial $\alpha$, since $U\cap D \in \mathbb Z$, the relevant gauge transformations are such that $u \notin \mathbb Z$; in other words, $\textrm{exp} (2 \pi u) \neq 1$, and the gauge transformations are not single-valued under $\theta \to \theta + 2\pi$. Such twisted gauge transformations can certainly be defined for a non-simply-connected gauge group like $U(1)$.

\bigskip\noindent{\it The Effective Field Strength In The Presence Of Surface Operators}

In any gauge theory, supersymmetric or not, the kinetic term of the gauge field has a positive-definite real part. As such, the Euclidean path-integral (which is what we will eventually be interested in) will be non-zero if and only if the contributions to the kinetic term  are strictly non-singular. Therefore, as a result of the singularity (\ref{surface operator}) when one includes a surface operator in the theory, the effective field strength in the Lagrangian that will contribute non-vanishingly to the path-integral must be a shifted version of the field strength $F$. In other words, whenever we have a surface operator along $D$, one ought to study the action with field strength $F' = F - 2\pi \alpha \delta_D$ instead of $F$. This means that the various fields of the theory are necessarily coupled to the gauge field $A'$  with field strength $F'$. This important fact was first pointed out in~\cite{Gukov-Witten}, and further exploited in~\cite{mine} to prove an $S$-duality in a general, abelian ${\CN} = 2$ theory without matter in the presence of surface operators.

\def\cq{{ \CQ}}

\def\mo{{\mathscr O}}

\newsubsection{A Physical Interpretation Of The ``Ramified'' Donaldson Invariants}

\medskip\noindent{\it Correlation Functions Of $\cq$-Invariant Observables}

\def\mI{{\mathscr I}}

Consider a topologically-twisted version of a pure $SU(2)$ or $SO(3)$ theory with $\CN =2$ supersymmetry -- also known as Donaldson-Witten theory -- in the presence of surface operators.  This theory has a nilpotent scalar supercharge $\cq$, and its action can be written as~\cite{u-plane SO}
\be
S_E =  { \{ \cq , V \}  \over e^2}  +   {i\Theta \over 8 \pi^2} \int_X \textrm{Tr} \hspace{0.1cm} F' \wedge F'  -  i  \int_X  \textrm{Tr}  \hspace{0.1cm} \eta  \delta_D \wedge F'
\label{S}
\ee
for some fermionic operator $V$ of $R$-charge -1 and scaling dimension 0, and complexified gauge coupling $\tau = {4 \pi i \over e^2} + {\Theta \over 2 \pi}$. The action is thus $\cq$-exact up to purely topological terms.  

Now consider the set of $\cq$-invariant observables ${\mathscr O}_i$ and their correlation function
\be
\langle \mo_1 \dots \mo_n \rangle = \int \CD \Phi \   \mo_1 \dots \mo_n  e^{- {S_E}},
\label{CF}
\ee
where $\CD \Phi$ denotes the total path-integral measure in all fields. Note that one of the central features of the twisted theory is that its stress tensor $T_{\mu \nu}$ is $\cq$-exact, i.e., $T_{\mu \nu} = \{ \cq, G_{\mu\nu}\}$ for some fermionic operator $G_{\mu\nu}$. Consequently, a variation of the correlation function with respect to the metric yields $\delta_{\bar g}\langle \mo_1 \dots \mo_n \rangle = -\langle \mo_1 \dots \mo_n \cdot {\delta S_E  \over \delta{\bar g}^{\mu \nu}} \rangle =  -\langle \mo_1 \dots \mo_n \cdot T_{\mu \nu} \rangle =  -\langle \mo_1 \dots \mo_n \cdot  \{ \cq, G_{\mu\nu}\} \rangle =  -\langle \{ \cq ,\mo_1 \dots \mo_n \, G_{\mu\nu}\} \rangle = 0$, where we have made use of the fact that $\langle \{ \cq, \dots \} \rangle = 0$ since $\cq$ generates a (super)symmetry of the theory. Notice also that a differentiation of the correlation function with respect to the gauge coupling $e$ yields
$
{\partial \over \partial e} \langle \mo_1 \dots \mo_n \rangle  = {2 \over e^3} \langle \mo_1 \dots \mo_n  \{ \cq , V \}  \rangle  = {2 \over e^3} \langle \{ \cq, \mo_1 \dots \mo_n \, V \}  \rangle = 0.
$
In other words, the correlation function of $\cq$-invariant observables is independent of the gauge coupling $e$; as such, the semiclassical approximation to its computation will be exact. In this approximation, one can freely send $e$ to a very small value in the correlation function. Consequently, from (\ref{S}) and (\ref{CF}), we see that the non-zero contributions to the correlation function will be centered around classical field configurations -- or the zero-modes of the fields -- which minimize $\{ \cq , V \}$ and therefore the action. Thus, it suffices to consider quadratic fluctuations around these zero-modes.

Let us first consider the fluctuations. Assuming that the operators $\mo_i$ can be expressed purely in terms of zero-modes, the path-integral over the fluctuations of the fields in the kinetic terms of the action give rise to determinants of the corresponding kinetic operators. Due to supersymmetry, the determinants resulting from the bose and fermi fields cancel up to a sign. (This point will be important when we physically derive Taubes' result in a later section).  

Let us now consider the zero-modes. The bosonic zero-modes obey the constraints obtained by setting the supersymmetric variation of the fermi fields in the twisted theory to zero. We find that these constraints are $F'_+ = 0$,  $\nabla' \phi =0$ and $[\phi, \phi^{\dagger}] =0$, where $\nabla'$ is the gauge-covariant derivative and $\phi$ is a complex bose field~\cite{u-plane SO}. If we assume the trivial solution $\phi =0$ to the constraints $\nabla' \phi =0$ and $[\phi, \phi^{\dagger}] =0$, it will mean that the zero-modes of $A'$  do $\textrm{\it not}$ correspond to reducible connections, and that there are $\it no$ zero-modes of $\phi$.  Altogether, this means that the only bosonic zero-modes come from the gauge field $A'$, and that they correspond to irreducible, anti-self-dual connections which are characterized by the relation
\be
F_+  = 2 \pi \alpha \delta^+_D.
\label{ram instantons}
\ee
Again, recall that the bundle $E'$  over $X$ with curvature $F' = F- 2\pi \alpha \delta_D$ can be equivalently viewed as the bundle $E$ with curvature $F$ restricted to $X \backslash D$. Hence, the constraint (\ref{ram instantons}) just defines  anti-self-dual connections on the bundle $E$ restricted to $X \backslash D$, whose holonomies around small circles linking $D$ are as given in (\ref{holonomy}). In other words, modulo gauge transformations that leave (\ref{ram instantons}) invariant, the expansion coefficients of the zero-modes of $A'$ that appear in the path-integral measure will correspond to the collective coordinates on $\CM'$ - the moduli space of ``ramified'' instantons.  

The rest of the fields in the theory are given by the fermions $(\zeta, \chi^+_{\mu\nu}, \psi_{\mu})$.  Since we have restricted ourselves to connections $A'$ that are irreducible, and moreover, if we assume that they are also regular, it can be argued that $\zeta$ and self-dual $\chi^+$ do not have any zero-modes~\cite{u-plane SO}. Thus, the only fermionic zero-modes whose expansion coefficients contribute to the path-integral measure come from $\psi$. 

The number of bosonic zero-modes is, according to our analysis above, given by the dimension $s$ of $\CM'$. What about the number of zero-modes of $\psi$? Well, since there are no zero-modes for $\zeta$ and $\chi^+$,  the dimension of the kernel of the kinetic operator $\Delta_F$ which acts on $\psi$ in the (``ramified'') Lagrangian is equal to the index of $\Delta_F$; in other word, the number of zero-modes of $\psi$ is given by ${\rm dim} ({\rm Ker} (\Delta_F)) = {\rm ind}(\Delta_F)$. This index also counts the number of infinitesimal connections $\delta A'$ where gauge-inequivalent classes of $A' + \delta A'$ satisfy $F'_+ = 0$, i.e.,  (\ref{ram instantons}). Therefore, the number of zero-modes of $\psi$ will also be given by the dimension $s$ of $\CM'$. Altogether, this means that after integrating out the non-zero modes,  we can write the remaining part of the measure in the expansion coefficients $a'_i$ and $\psi_i$ of the zero-modes of $A'$ and $\psi$ as
\be
\Pi^s_{i=1} da'_i d\psi_i.
\label{measure} 
\ee
Notice that the $s$ distinct $d\psi_i$'s anti-commute. Hence, (\ref{measure}) can be interpreted as a natural measure for the integration of a differential form in $\CM'$.

\newpage
\bigskip\noindent{\it Correlation Functions Corresponding To The ``Ramified'' Donaldson Invariants} 

In the relevant case that $b_1(X) =0$, one has the following $\cq$-invariant observables 
\begin{eqnarray}
\label{I's zero-modes 1}
I'_0(p) & = &{1 \over 8 \pi^2}\textrm{Tr}{ \langle \phi (p) \rangle}^2   \\ 
I'_2(S) & = & -{1 \over {\sqrt {32}} \pi^2} \int_S  \textrm{Tr} \left (\langle \phi \rangle F \right)_0
\label{I's zero-modes}
\end{eqnarray}
for any $p \in H_0(X \backslash D)$ and $S \in H_2(X \backslash D)$; up to lowest order in $e$ in the semiclassical approximation, 
\be
\langle \phi (x) \rangle = - {i \over \sqrt 2} \int_X d^4y \,  G(x-y)[\psi(x), \psi(y)]_0, 
\label{<phi>}
\ee
where $G(x-y)$ is the unique solution to the relation $\nabla'^2 G(x-y) = \delta^4(x-y)$;  the subscript ``0'' in (\ref{I's zero-modes}) and (\ref{<phi>}) just denotes their restriction to zero-modes.

Like the assumption made of the operators $\mo_i$ in (\ref{CF}), $I'_0(p)$ and $I'_2(S)$ are express purely in terms of zero-modes. Moreover, based on our above discussion about  (\ref{measure}) being a natural measure for the integration of differential forms in $\CM'$, we see that $I'_0(p)$ and $I'_2(S)$ (which contain 4 and 2 zero-modes of $\psi$, respectively) can be interpreted as 4-forms and 2-forms in $\CM'$. 

Let us compute an arbitrary correlation function in the $\cq$-invariant observables $I'_0(p)$ and $I'_2(S)$. For the correlation function to be non-vanishing, the $d\psi_i$'s in the remaining measure (\ref{measure})  have to be  ``soaked up'' by the  zero-modes of $\psi$  that appear in the combined operator whose correlation function we wish to consider, $\it exactly$. This just reflects the fact that a non-vanishing correlation function is necessarily $R$-invariant: the field $\psi$ carries a non-zero $R$-charge of $1$, and under an $R$-transformation, the integration measure and an appropriately chosen combined operator will transform with weights $-\Delta R$ and  $\Delta R$, respectively, where $\Delta R$ is the number $N_{\psi}$ of zero-modes of $\psi$. In turn, this means that the combined operator ought to correspond to a top-form (of degree $s$) in $\CM'$. Therefore, if such a combined operator is given by $[{I'_0}(p)]^t I'_2(S_{i_1}) \dots I'_2(S_{i_m})$,\footnote{Notice that we are considering a combined operator in which there are $t$ operators $I'_0(p_{i_1}), \dots, I'_0(p_{i_t})$ that coincide at one particular point $p$ in $X$. Such a combined operator can be consistently defined in any physical correlation function; this is because the  $I'_0(p_{i_k})$'s consist only of non-interacting zero-modes and moreover, any correlation function of the topological theory is itself independent of the insertion points of the operators.} its $\it topological$ correlation function -- for ``$\it ramified$'' instanton number $k' = -  \int_X p_1(E') / 4$ -- can be written as
\be
\langle  [{I'_0}(p)]^t I'_2(S_{i_1}) \dots I'_2(S_{i_m})   \rangle_{k'} = \int_{\CM'}  [{I'_0}(p)]^t \wedge I'_2(S_{i_1}) \wedge \dots \wedge 
I'_2(S_{i_m}),
\label{correlation function 1}
\ee
where $2m + 4t = s$. This coincides with the definition of the ``ramified'' Donaldson invariants $d^{k'}_{m,t}$ in (\ref{ram DW}). Thus, we have found, in  (\ref{correlation function 1}),  a physical interpretation of the ``ramified'' Donaldson invariants in terms of the correlation functions of $\cq$-invariants observables $I'_0(p)$ and $I'_2(S)$. As a result, the generating function ${\bf Z}'_{\xi, \bar g}(p,S)$ in (\ref{Donaldson generating function}) can also be interpreted in terms of $I'_0$ and $I'_2$ as 
\be
{\bf Z}'_{\xi, \bar g}(p,S) =  \sum_{k'} \  \langle e^{p I'_0 +  I'_2(S)} \rangle_{k'}.
\label{Donaldson generating function - physical}
\ee
The ``ramified'' Donaldson series in (\ref{series}) is then given by
\be
\label{series physical}
{\mathscr D}' _\xi (S) = \sum_{k'} \, \left(1 + {1\over 2} {\partial \over \partial p}\right) \cdot \langle e^{p I'_0 +  I'_2(S)} \rangle_{k'} \vert_{p=0}. 
\ee

\newsubsection{The ``Ramified'' Seiberg-Witten Equations And Invariants}

In the topologically-twisted version of the corresponding low-energy Seiberg-Witten (SW) theory with surface operators, the supersymmetric configurations correspond to solutions of the equations~\cite{u-plane SO}
\be
(F^d_L)_+ = (\overline M M)_+  +  2 \pi \alpha_d \delta^+_D
\label{SW1}
\ee  
and 
\be
\dirac M =0,
\label{SW2}
\ee
where $\dirac$ is the Dirac operator $\it coupled$ to the effective $U(1)$ photon with field strength ${F^d_L}' = F^d_L - 2 \pi \alpha_d \delta_D$; the label ``$d$'' indicates that the field or parameter is that which is defined in the preferred \emph{dual} ``magnetic'' frame;  and $M$ is a section of the complex vector bundle $S_+ \otimes L'_d$, where $S_+$ and $L'_d$ are a positive spinor bundle and a $U(1)$-bundle with curvature field strength ${F^d_L}'$, respectively. (\ref{SW1}) and (\ref{SW2}) together define the ``ramified'' SW equations, whence the relevant $\it topological$ correlation functions in the case where $b_1(X) =0$ are
\be
 \langle  [{J^d_0}(p)]^{q}  \rangle_{\lambda'} = \int_{\CM^{\lambda'}_{\textrm{sw}}} [{J^d_0}(p)]^{q} = SW_{\lambda'}.
\label{correlation function SW}
\ee
Here, $q = d^{\lambda'}_{\textrm {sw}} /2$, where  $d^{\lambda'}_{\textrm{sw}}$ -- the even dimension of the moduli space $\CM^{\lambda'}_{\textrm{sw}}$ of the ``ramified'' SW equations determined by the ``ramified'' first Chern class $\lambda' =  {1\over 2}c_1({L'_d}^{\otimes 2})$ of the determinant line bundle ${L'_d}^{\otimes 2}$ of the $\spin$-structure associated with a choice of  the complex vector bundle $S_+ \otimes L'_d$ -- is given by~\cite{Appendix 1}
\be
d^{\lambda'}_{\textrm{sw}} =  - {2\chi + 3 \sigma \over 4} + (\lambda')^2.
\ee
Also 
\be
\label{a_d}
{J^d_0}(p) = \langle \varphi_d(p) \rangle = a_d,
\ee 
where $p \in H_0(X)$,\footnote{Note that in contrast to the physical definition of the correlation functions which correspond to the ``ramified'' Donaldson invariants, here, we need not restricted the zero-cycles $p$ to $X \backslash D$. This is because the operator $a_d$ -- unlike $F$, $F_L$ or $F^d_L$ -- does not contain a singularity along $D$.} and $\varphi_d$ is a complex scalar in the ``magnetic'' $\CN=2$ vector multiplet of the SW theory. As required, ${J^d_0}(p)$ is expressed purely in terms of non-fluctuating zero-modes; it has $R$-charge 2 (associated with an accidental $U(1)_R$ symmetry at low-energy) and consequently, it can be interpreted as a 2-form in $\CM^{\lambda'}_{\textrm{sw}}$. Following~\cite{u-plane SO}, let us call $SW_{\lambda'}$ the ``ramified'' SW invariant for the basic class $\lambda'$.

\bigskip\noindent{\it When $X$ Is Of  (``Ramified'') Seiberg-Witten Simple-Type}

If $X$ is of ``ramified'' SW simple-type, i.e., ${\rm dim} (\CM^{\lambda'}_{\textrm{sw}}) = 0$, we have $q=0$ in (\ref{correlation function SW}), and as explained in~\cite{Appendix 1}, we have
\be
SW_{\lambda'} = SW(\lambda') = \sum_{x_i} \, (-1)^{n_{i}},
\label{exp SW form}
\ee 
where the $x_i$'s are the points that span the zero-dimensional space $\CM^{\lambda'}_{\textrm{sw}}$, and the $n_i$'s are integers which are determined by the corresponding sign of the determinant of an elliptic operator associated with a linearization of the ``ramified'' SW equations. In other words, $SW(\lambda')$ counts (with signs) the number of solutions of the ``ramified'' SW equations determined by $\lambda'$; in particular, it is an \emph{integer}, just like its ordinary counterpart. 

Recall that the ordinary limit whence there is effectively no ``ramification'' along $D$ is achieved when the effective value of $\alpha_d$ approaches an integer. Since the foregoing discussion holds for  arbitrary values of $\alpha_d$, a four-manifold of ``ramified'' SW simple-type is necessarily of ordinary SW simple-type, too.

\newsection{Physical Proofs Of Seminal Theorems By Ozsv\'ath And Szab\'o}

In this section, physical proofs of various seminal theorems in four-dimensional geometric topology obtained by Ozsv\'ath and Szab\'o in~\cite{OS, OS 2}, will be furnished. Our computations in this section will soon prove to be useful when we physically derive Taubes' spectacular result  in the next section. 

\newsubsection{Adjunction Inequality For Embedded Surfaces Of Negative Self-Intersection}

In Corollary 1.7 of their seminal paper~\cite{OS}, Ozsv\'ath and Szab\'o demonstrated that embedded surfaces of \emph{negative} self-intersection in four-manifolds actually obey an adjunction inequality which involves the first Chern class of the $\textrm{Spin}^c$-structure. We will now present a physical proof of this mathematical corollary. 

\bigskip\noindent{\it A Minimal Genus Formula From $R$-Invariance}

Firstly, a relevant result from~\cite{u-plane SO} is the following. Consider $X$ with $b_1=0$ and odd $b^+_2 > 1$; assume that $g \geq 1$, where $g$ is the genus of the surface operator $D \subset X$; then, for \emph{any} $D\cap D \neq 0$, $R$-invariance of the non-vanishing correlation functions in (\ref{correlation function 1}) will imply that
\be
 2 D \cap D - (2g -2) \leq  4l \leq (2g -2),
 \label{inequality}
\ee
where $l = \int_D F_L / 2\pi$. 

Secondly, since $g \geq 1$ and therefore $(2g -2) \geq 0$, we can infer from (\ref{inequality}) that
\be
  (2g -2) \geq D \cap D   - 2 l.
 \label{KM's result}
\ee
Moreover, note that due to electric-magnetic duality, $F_L$ is physically $\it equivalent$ to the field strength $F^d_L$ of the low-energy $U(1)$ theory; in turn, $F^d_L$ corresponds to $- \pi c_1(L^2_d)$. In other words, we can identify  $-2l$  with $c_1(L^2_d)[D]$.  Consequently, we can write (\ref{KM's result}) as
\be
(2g -2) \geq D \cap D   + c_1(L^2_d)[D].
\label{result now}
\ee
The above formula coincides with Theorem 1.7(b) of~\cite{structure}.

\bigskip\noindent{\it And The Proof}

Now, let us consider a surface operator $D = \Sigma$ with $\Sigma \cap \Sigma \leq 0$ and genus $g \geq 1$. Since (\ref{inequality}) is valid for any value of $D \cap D$, we can write 
\be
 2  \Sigma \cap \Sigma - (2g -2) \leq  4l \leq (2g -2).
 \label{inequality OS}
\ee
 Since $(2g -2) \geq 0$ and  $\Sigma \cap \Sigma \leq 0$, we obtain from (\ref{inequality OS}) the following inequality 
\be
(2g-2) \geq  \Sigma \cap \Sigma - c_1(L^2_d)[\Sigma],
\label{ineq above}
\ee
after identifying $-2l$ with $c_1(L^2_d)[\Sigma]$. 

As (\ref{result now}) is also valid for arbitrary values of $D \cap D$, it will mean that
\be
(2g -2) \geq \Sigma \cap \Sigma   + c_1(L^2_d)[\Sigma].
\label{result now 1}
\ee
Thus, if (\ref{ineq above}) and (\ref{result now 1}) are to hold simultaneously, it will mean that
\be
\vert  c_1(L_d^2)[\Sigma] \vert  + \Sigma \cap \Sigma \leq (2g-2).
\label{cor 1.7}
\ee
This is just Corollary 1.7 of~\cite{OS}. In fact, our physical proof asserts that (\ref{cor 1.7}) should also hold for $\Sigma$ with $\Sigma \cap \Sigma = 0$, and not just for $\Sigma$ with $\Sigma \cap \Sigma < 0$ (as stipulated in Corollary 1.7 of~\cite{OS}); this physical assertion is indeed consistent with Theorem 1.1 of~\cite{OS 2} for $X$ of SW  simple-type (for which (\ref{cor 1.7}) is also valid). 

 \newsubsection{A Relation Among The Ordinary Seiberg-Witten Invariants}

Ozsv\'ath and Szab\'o also showed in Theorem 1.3 of~\cite{OS} and Theorem 1.6 of~\cite{OS 2}, that there exists relations among the ordinary SW invariants which arise from the above embedded surfaces with negative self-intersection in $X$ with $b^+_2(X) > 1$. We will now present the physical proofs of these mathematical theorems.

\bigskip\noindent{\it The ``Magic'' Formula}

\def\tD{{\widetilde \Sigma}}

\def\lamba{{\bar \lambda}}

First, note that for $X$ with $b_1=0$ and $b^+_2 > 1$,  the ``magic'' formula which expresses the generating function $Z'_{\rm D}$ of the ``ramified'' Donaldson invariants in terms of the (``ramified'') SW invariants when $\Sigma \cap \Sigma \neq 0$, is (via (7.20) and (7.15) of~\cite{u-plane SO})
\begin{eqnarray}
\label{magic}
Z'_{\rm D} & = &  \sum_{\lamba} \, {SW_\lamba \over 16} \cdot e^{2i\pi(\lambda_0\cdot\lamba+\lambda_0^2)} \cdot e^{2\lamba[\tD]} \nonumber \\
&& \cdot {\rm Res}_{q_M=0} \left[{dq_M\over q_M} q_M^{-\lamba^2/2} {\vartheta_2^{8+\sigma} \over  a_d h_M } \left(2i {a_d \over  h_M^2 }\right)^{(\chi+\sigma)/4} \exp\left[ 2 p u_M - i(\lamba,S)/h_M + S^2 T^M_S \right] \right] \nonumber \\
&& + i^{\{(\chi + \sigma) / 4  -w_2(E)^2\}} \sum_{\lambda'} \, {SW_{\lambda'} \over 16} \cdot e^{2i\pi(\lambda_0\cdot\lambda+\lambda_0^2 + \alpha^2 \Sigma^2 /2)}  \cdot e^{2\lambda[\tD]}  \\
&& \hspace{-0.7cm} \cdot {\rm Res}_{q_M=0} \left[{dq_M\over q_M} q_M^{-(\lambda')^2/2} {\vartheta_2^{8+\sigma} \over  a_d h_M } \left(2i {a_d \over  h_M^2 }\right)^{(\chi+\sigma)/4} \exp\left[ -2 p u_M + i(\lambda, i S)/h_M  - S^2 T^M_S - 4 \tD^2 T^M_{\tD} \right] \right],\nonumber  
\end{eqnarray}
where $\lambda' = \lambda - \alpha\delta_\Sigma$ is a ``\emph{ramified}'' (first Chern class of the) $\spin$-structure for effectively nontrivial values of $\alpha$; $\vartheta_2^{8+\sigma}(\tau)$ is a certain Jacobi theta function in $q_M = e^{2\pi i \tau}$, while $a_d$, $u_M$, $h_M$, $T^M_S$ and $T^M_{\tD}$ are polynomial functions in $q_M$ (see appendix A and $\S$4.2 of~\cite{u-plane SO} for their explicit expansions); $\tD = i \pi \alpha \Sigma /2$; $2\lambda_0$ is an integral lift of $w_2(E)$; and $\lamba$ is an \emph{ordinary} (first Chern class of the) $\spin$-structure.

Second, let us specialize to the case where the microscopic gauge group is $SU(2)$; i.e., $\xi = w_2(E) = 0$. In this case, $\lambda_0$ can be set to zero in the ``ramified'' and ordinary theories~\cite{u-plane SO, Moore-Witten}. Moreover, if we assume $X$ to be such that the values of $\lamba^2 /2$ and $(\lambda')^2/2$ of the first and second residues in~(\ref{magic}), respectively, are both given by $(\chi + \sigma)/4 + \sigma/8$,  the computation of (\ref{magic}) will simplify considerably; only the leading terms in the $q_M$-expansion contribute non-vanishingly. Using $u_M = 1 + \dots$, $T^M_S = 1/2 + \dots$, $T^M_\tD =  1/4 + \dots$,  $h_M = 1/ (2i) + \dots$ and $a_d = 16 i q_M + \dots$,  one will compute (\ref{magic})  to be
\be
\label{Zd for OS proof simplified}
Z'_{\rm D}  =  2^{1 + {7\chi \over 4} + {11 \sigma \over 4} } \left \{ \sum_{\lamba} \, {SW_\lamba} \, e^{2 p  + S^2 /2} \, e^{2(S + \tD, \lamba)}  + i^{(\chi + \sigma) / 4  } \sum_{\lambda'} \, {SW_{\lambda'}} \, (-1)^{\alpha^2 \Sigma^2} \, e^{-2 p  - S ^2 /2 -  \tD^2} \, e^{-2 i (S + i \tD, \lambda)} \right  \}. 
\ee

\bigskip\noindent{\it On To The Proofs}

\def\half{{1 \over 2}}

\def\lamhat{\hat \lambda}

Before we proceed further, note that since our objective is to provide a physical proof of a mathematical result, one needs to express (\ref{Zd for OS proof simplified}) in the mathematical convention. One can do so by replacing $\alpha$ with $i \alpha$ and the relevant $U(1)$ field strengths $F$ with $iF$, throughout; the gauge fields are then valued in the complex Lie algebra, as desired. 

Coming back to our main discussion, let us send the effective value of $\alpha$ to $\pm 1$.  Then, the condition $(\lambda - i\alpha \delta_\Sigma)^2 /2 = (\chi + \sigma)/4 + \sigma/8$ at the dyon point (i.e., the second contribution in $Z'_{\rm D}$) implies that  we have $d_{L^2_d} = \lambda^2 - (2 \chi + 3 \sigma) /4 = \mp c_1(L_d^2)[\Sigma] +  \Sigma \cap \Sigma$. As there is effectively no  ``ramification'' in the $U(1)$ theory at the dyon point when $\alpha = \pm 1$, we can, via (\ref{correlation function SW}) and (\ref{a_d}), write $SW_{\lambda'}$ in (\ref{Zd for OS proof simplified}) as
\be
SW_{\lambda'} = \int_{{\cal M}^{\lambda'}_{\rm sw}} (a_d)^{d_{L^2_d}/2},
\label{exp SW}
\ee 
where ${\cal M}^{\lambda'}_{\rm sw}$ is the moduli space of the ordinary SW equations whose even dimension is therefore $d_{L^2_d}$. 
As explained in footnote~6, $a_d$ can be defined at \emph{any} point in $X$; thus, let us, for later convenience, define $a_d$ at some point $z \in \Sigma$.

When $\alpha = \pm 1$, there is also no ``ramification'' in the microscopic $SU(2)$ theory associated with $Z'_{\rm D}$ on the LHS of (\ref{Zd for OS proof simplified}); moreover, recall from $\S$2.2 that modular-invariance requires that the phase term (\ref{eta term}) be equal to $1$ whenever the surface operators are nontrivially-embedded; as such, one can express $Z'_{\rm D}$ on the LHS of (\ref{Zd for OS proof simplified}) as eqn.~(7.25) of~\cite{u-plane SO} via Witten's ordinary magic formula. Furthermore, since the condition $\lamba^2 /2 = (\chi + \sigma)/4 + \sigma/8$ implies that $X$ is of SW simple-type, one can denote $SW_\lamba$ as $SW(\lamba)$ on the RHS of (\ref{Zd for OS proof simplified}). Last but not least, note that one can appeal to a regular gauge transformation of the kind in~(\ref{twisted gauge tx}) with $u = \mp 1$ -- \emph{which leaves the gauge-invariant $\lambda'$ unchanged} -- to shift $\alpha$ and therefore $\tD$ to zero on the RHS of (\ref{Zd for OS proof simplified}). Altogether, this means that one can also express (\ref{Zd for OS proof simplified}) as  
\begin{eqnarray}
\label{OS proof compare}
 \sum_{{\hat \lambda}} \left \{ {SW({\hat \lambda})} \,  e^{2 p  + S^2 /2} e^{2(S, {\hat \lambda})}  + i^{(\chi + \sigma) / 4}  \,{SW({\hat \lambda})} \, e^{-2 p  - S^2 /2} e^{-2i(S, {\hat \lambda})} \right \} \nonumber \\
 =   \sum_{\lambda} \left \{ SW(\lambda) \,  e^{2 p  + S^2 /2} \, e^{2(S, \lambda)}  + i^{(\chi + \sigma) / 4} \, {SW_{\lambda'}}  \,  e^{-2 p  - S^2 /2} \, e^{-2 i (S, \lambda)} \right \}. \nonumber \\ 
\end{eqnarray}
In (\ref{OS proof compare}), $\hat \lambda$ is an ordinary $\spin$-structure; $\lambda' = \lambda \mp i \delta_\Sigma$; and $\lambda = -F^d_L / 2\pi$, where $F^d_L$ is an ordinary $U(1)$ field strength, i.e., the holonomy of its gauge field around a small circle that links $\Sigma$ is trivial. Via a term-by-term comparison of (\ref{OS proof compare}), we conclude that we have an equivalence
\be
SW_{{\frak s}'} = SW(\frak s)
\label{SW = SW}
\ee
of \emph{ordinary} SW invariants, where ${\frak s}' = -i \lambda'$ and $\frak s = -i \lambda$ are the respective (first Chern class of the) $\spin$-structures expressed in the mathematical convention.

A few observations are in order. First, notice that if $d_{L^2_d} = \vert c_1(L_d^2)[\Sigma] \vert +  \Sigma \cap \Sigma$, then ${\frak s}' = {\frak s} + \epsilon \delta_\Sigma$, where $\epsilon = \pm 1$ is the sign of $c_1(L_d^2)[\Sigma]$. Second, since the scalar variable $a_d$ has $R$-charge $2$, it will represent a class in $H^2({\cal M}^{{\frak s}'}_{\rm sw})$. Third, because we have assumed that $b_1(X) =0$, it will mean that $H_1(X, \mathbb Z)$ is empty. Fourth,  as $d_{L^2_d} \geq 0$,  it will mean that $ \vert c_1(L_d^2)[\Sigma] \vert + \Sigma \cap \Sigma \geq 0$; together with (\ref{cor 1.7}), this implies that $g > 0$. With these four points in mind, it is thus clear that (\ref{exp SW}) and (\ref{SW = SW}) are  \emph{precisely} Theorem 1.3 of~\cite{OS}; here, $a_d$ and $d_{L^2_d} /2$ can be  identified  with $U$ and $(m+ g)$ in Theorem 1.3 of~\cite{OS},  respectively, while $a$  in Theorem 1.3 of~\cite{OS} can be set to 1 since $X$  is of SW simple-type. 

\bigskip\noindent{\it When $X$ Is Not Of SW Simple-Type}

What if $X$ is not of SW simple-type? The analysis is similar: one just substitutes $\lambda^2 /2$ and $(\lambda')^2/2$ as $(\chi + \sigma)/4 + \sigma/8 + p$ -- where $p$ is some fixed positive integer -- in the first and second residues of~(\ref{magic}), respectively, and proceed as above. The only difference now is that the effective value of $d_{L^2_d}$ will be shifted by $2p$, and instead of (\ref{SW = SW}), we will have 
\be
SW_{\frak s'} = SW_{\frak s},
\label{final OS}
\ee
where
\be
SW_{\frak s'} = \int_{{\cal M}^{\frak s'}_{\rm sw}} (a_d)^{d_{L^2_d}/2} \, (a_d)^p \, \qquad {\rm and} \, \qquad SW_{\frak s} = \int_{{\cal M}^{\frak s}_{\rm sw}} (a_d)^p.
\label{SW int exp}
\ee
In this case, $a$ of~\cite{OS} is no longer equal to 1 but rather, it is $U^p \in \mathbb A(X)$, where $\mathbb A (X)$ is the polynomial algebra $\mathbb Z[U]$.  It is also clear from (\ref{SW int exp}) that $2p$ must be equal to the dimension of the moduli space ${\cal M}^{\frak s}_{\rm sw}$ of the SW equations with $\spin$-structure $\frak s$. 

\bigskip\noindent{\it When $b^+_2 (X)  =1$}

And what if $b^+_2 =1$? In this case, as explained in $\S$6.3 of~\cite{u-plane SO}, the monopole and  dyon point contributions to $Z'_D$  -- which depend on $SW_\lambda$ and $SW_{\lama}$, respectively -- will jump as we cross certain ``walls'' in the forward light cone $V_+= \{\omega_+ \in H^{2,+}(X;\IR): (\omega_+)^2 > 0 \}$. In particular, $SW_\lambda$ will jump if we cross the ``wall'' defined by 
\be
(\omega_+, \lambda)  =  {i \alpha_M \over 2} (\omega_+, \Sigma), 
\label{wall M}
\ee
where $\alpha_M = - \eta$, while $SW_\lama$ will jump if we cross the ``wall'' defined by
\be
(\omega_+, \lambda)  =  {i \alpha_D \over 2} (\omega_+, \Sigma),
\label{wall D}
\ee
where $\alpha_D = 2 \alpha$. 

Notice that when $\alpha = \pm 1$, we can rewrite (\ref{wall D}) as
\be
(\omega_+, \frak s + \epsilon \delta_\Sigma) = 0,
\label{wall d}
\ee
where $\epsilon = \mp$. 

Recall also that since modular-invariance requires $\eta$ to vanish whenever we have a nontrivially-embedded surface operator such as $\Sigma$, the RHS of (\ref{wall M}) is identically zero. This is tantamount to setting 
\be
(\omega_+, \Sigma) = 0. 
\label{PD}
\ee
Consequently, one can rewrite (\ref{wall M}) as
\be
(\omega_+, \frak s) = 0. 
\label{wall m}
\ee

Therefore, for (\ref{SW = SW}) or (\ref{final OS}) to continue to hold unambiguously when $b^+_2 =1$,  $\omega_+$ must not lie anywhere along the ``walls'' defined by (\ref{wall d}) and (\ref{wall m}) where the values of the LHS and RHS of (\ref{SW = SW}) or (\ref{final OS}), respectively, will jump; in addition, $\omega_+$ must also satisfy (\ref{PD}). This observation matches \emph{exactly} the claim in Theorem 1.3 of~\cite{OS} for four-manifolds with $b_2^+ =1$. This completes our physical proof of Theorem 1.3 of~\cite{OS}. 

\bigskip\noindent{\it When $\Sigma$ Has Arbitrary Self-Intersection Number}

Finally, note that  (\ref{SW = SW}) and its generalization (\ref{final OS}) also hold for $\Sigma$ with \emph{arbitrary} (as opposed to just negative) self-intersection number; this has been proved mathematically as Theorem 1.6 of~\cite{OS 2} by Ozsv\'ath and Szab\'o.  Once more, one can furnish a physical proof of this theorem.  
\def\dld {{d_{L^2_d}}}

To this end, let us consider the case where $\alpha = - \epsilon = 1$. Then, $d_{L^2_d} = - c_1(L_d^2)[\Sigma] +  \Sigma \cap \Sigma$. Since $d_{L^2_d} \geq 0$, we have 
\be
- c_1(L_d^2)[\Sigma] +  \Sigma \cap \Sigma \geq 0.
\label{in 1}
\ee

On the other hand, since $\dld$ can be identified with $2(m+g)$ in~\cite{OS} (and therefore~\cite{OS 2}), as $2m \geq 0$, it will mean that $\dld -2g \geq 0$. In turn, this implies that
\be
- c_1(L_d^2)[\Sigma] +  \Sigma \cap \Sigma + 2p > 2g -2,
\label{in 2}
\ee 
since $p$ is a positive integer. 

As we have not appealed to (\ref{cor 1.7}) in deducing the above inequalities, the self-intersection number of $\Sigma$ is allowed to be arbitrary in (\ref{in 1}) and (\ref{in 2}). Consequently, after noting that $\frak s' = \frak s - \delta_\Sigma$ because $\epsilon =-1$, we find that (\ref{in 1}) and (\ref{in 2}), with (\ref{final OS}) and (\ref{SW int exp}),  is nothing but Theorem 1.6 of~\cite{OS 2}; here, $a_d$, $p$ and $(\dld/2 + p)$ can be consistently identified as $U$, $d$ and $d'$ in Theorem 1.6 of~\cite{OS 2}, respectively. When $b^+_2(X) =1$, these relations will continue to hold as long as the above-stated conditions on $\omega_+$ are satisfied.

\newsection{A Physical Derivation Of Taubes' Groundbreaking Result}

In a series of four long papers collected in~\cite{CT 1}, C.H.~Taubes showed that on any compact, oriented symplectic four-manifold $X$ with $b^+_2 > 1$, the ordinary SW invariants are (up to a sign) equal to what is now known as the Gromov-Taubes invariants which count (with signs) the number of pseudo-holomorphic (complex) curves which can be embedded in $X$. This astonishing result, as formidable as its mathematical proof may be, nonetheless lends itself to a simple and concrete physical derivation, as we shall now demonstrate.   

\newpage
\bigskip\noindent{\it  Pseudo-Holomorphic Curves In A Symplectic Four-Manifold}

\def\osp{{\omega_{\rm sp}}}

Let $\omega_{\rm sp}$ be a self-dual symplectic two-form  on $X$ that is compatible with an almost-complex structure $J$. Let $K$ be the canonical line bundle on $X$. If a surface operator $\Sigma$ is a\emph{ pseudo-holomorphic} curve embedded in $X$ (in the sense of Gromov~\cite{Gromov}), $J$ will map the tangent space of $\Sigma$ to itself.  Moreover, we will have $\int_{\Sigma} \osp > 0$; in other words, $\Sigma$ will be homologically nontrivial so that the Poincar\'e dual $\delta_\Sigma$ of its fundamental class lies in $H^2(X ,\mathbb Z)$. 

A connected $\Sigma$ is also known to satisfy the adjunction formula
\be
2 -2g + \Sigma \cap \Sigma = - c_1(K) [\Sigma].  
\label{adj Taubes}
\ee 
This implies that a flat torus with zero self-intersection number can potentially have multiplicity greater than 1; consequently, counting of such curves can be a delicate issue~\cite{CT 1}. Therefore, for simplicity, let us choose $\Sigma$ to be curved with a \emph{non-zero} self-intersection number. Such a choice is guaranteed by the fact that one can always find a basis of homology two-cycles $\{U_i\}_{i =1, \dots, b_2}$ in $X$ that has a purely diagonal, unimodular intersection matrix, whence $\Sigma$ can a priori be selected from the $b_2$ number of $U_i$'s with $\Sigma \cap \Sigma \neq 0$. 

However, since being compatible with $J$ implies that $\osp \in H^{2,+}(X, \mathbb R)$, together with $(\osp, \Sigma) > 0$, it will mean that there can be at most  $b^+_2$ choices of $\Sigma$ among the $U_i$'s such that
\be
 \delta_{\Sigma} = \delta^+_{\Sigma}
\label{PD sigma}
\ee
for some connected, non-multiply-covered, pseudo-holomorphic curve $\Sigma \subset X$ which obeys $\Sigma \cap \Sigma > 0$. (Exceptional spheres which may be multiply-covered are also being automatically excluded here since they have negative self-intersections~\cite{dusa}.)

\bigskip\noindent{\it A Particular  Instanton Sector}

\def\cl{{\cal L}}

As in the previous section, let us now consider the case where $E$ can be lifted to an $SU(2)$-bundle, i.e., $w_2(E) = \xi =0$. At low energies, the $SU(2)$ gauge symmetry is spontaneously-broken to a $U(1)$ gauge symmetry in the underlying physical theory. Mathematically, this means that $E$ can be expressed over all of $X$ as
\be
E = {L} \oplus {L}^{-1}
\ee 
at macroscopic scales, where $L$ is the complex line bundle corresponding to the unbroken $U(1)$ gauge symmetry. This implies that 
\be
c_2(E) =  - c_1(L)^2. 
\label{c=c}
\ee 
However, since we have an equivalence of characteristic classes which are themselves topological invariants,  it will mean that (\ref{c=c}) will also hold in the microscopic $SU(2)$ theory; in particular, since $k$ and $l$ are given by $\int_X c_2(E)$ and $-\int_\Sigma c_1(L)$, respectively, the values of $k$ and $l$ will be correlated for any particular choice of $X$ and surface operator $\Sigma$. 

Now consider the sector of the $SU(2)$ theory where $N_{\psi}$ is \emph{zero}; this is the sector where\footnote{Based on our discussions in $\S$2.2, the expression for the index of the kinetic operator $\Delta_F$ of $\psi$  that counts the number $N_{\psi}$ of $\psi$ zero-modes, is the expression for the index in the ordinary case but with gauge bundle $E'$. In other words, $N_{\psi} = 8k' - {3 \over 2} (\chi + \sigma)$, where the ``ramified'' instanton number $k' = - {1\over 8 \pi^2} \int_X \textrm{Tr} F' \wedge F'$.} 
\be
k' = {3 \over 16} (\chi + \sigma).
\label{k'}
\ee
If ${\bf Z}^{p'}_{0, \bar g}(p,S)$ is the $p'$-instanton sector  of  the generating function (\ref{Donaldson generating function - physical}) of the ``ramified'' $SU(2)$ Donaldson invariants of $X$, then
\be
  {\bf Z}^{k'}_{0, \bar g}(p, S) = {\bf Z}^{k'}_{0, \bar g}(0, 0) =    \langle 1 \rangle_{k'}.
\label{0-instanton}
\ee
Let ${\mathscr D}^{p'}_{0} (S)$ be the $p'$-instanton sector of the ``ramified'' $SU(2)$ Donaldson series ${\mathscr D}'_{0} (S)$ in (\ref{series physical}); then, from (\ref{0-instanton}) and (\ref{series physical}), we have  
\be
{\mathscr D}^{k'}_0 (S) = {\mathscr D}^{k'}_0 (0)  =   \langle 1 \rangle_{k'}.
\label{D_0}
\ee

At any rate, note that if $X$ is of (``ramified'') SW simple-type, from (\ref{Zd for OS proof simplified}), we have 
\be
\sum_{p'} {\mathscr D}^{p'}_{0} (S) =  \sum_{\lamba} \, SW(\lamba)  \, e^{2(S + \tD, \lamba) +  S^2 /2 + f(\chi + \sigma)},
\label{series simple-type}
\ee
where $f(\chi + \sigma)$ is a real-valued function in $\chi$ and $\sigma$. Two conclusions can be drawn  from (\ref{series simple-type}) at this point. First, since there is, as explained in $\S$3.1, a one-to-one correspondence between $l$ and $ -\int_{\Sigma} F^d_L /2 \pi$ due to electric-magnetic duality in the low-energy $U(1)$ theory, it will mean -- via the relation $k' = k + 2\alpha l - \alpha^2 \Sigma \cap \Sigma$,\footnote{Recall here that  $k ' = - {1\over 8 \pi^2} \int_X \textrm{Tr} F' \wedge F' = k +  \textrm{Tr} \alpha {\frak m} -  {1\over 2} \textrm{Tr} \alpha^2 \Sigma \cap \Sigma$, and since $\frak t$ is generated by (\ref{frak t}), it will mean that $k' =  k +  2 \alpha {l} -  \alpha^2 \Sigma \cap \Sigma$.}  and the correlation between $k$ and $l$ for any particular choice of $X$, $\Sigma \cap \Sigma$ and $\alpha$ -- that there is also a one-to-one correspondence between $k'$ and a \emph{certain} basic class $\lambda$.\footnote{Recall here that $\lambda = -F^d_L /2 \pi$. In fact, there is a one-to-one correspondence between all values of $p'$ and $\lamba$ in (\ref{series simple-type}), although the RHS of (\ref{series simple-type}) is known to consist of a finite number of terms only because some of the $SW(\lamba)$'s are zero: cancellations can occur in the $SW(\lamba)$'s since they take the form given in (\ref{exp SW form}).} Hence, ${\mathscr D}^{k'}_{0} (0)$ on the LHS of  (\ref{series simple-type}) will correspond to the $\lambda$-term on the RHS (\ref{series simple-type}). Second, as our notation indicates, ${\mathscr D}^{k'}_{0} (0)$ is independent of  $S$; also, in a supersymmetric topological quantum field theory whence the semiclassical approximation is exact, (\ref{D_0}) will mean that the topological invariant ${\mathscr D}^{k'}_{0} (0)$ -- like $SW(\lambda)$ of the $\lambda$-term on the RHS of (\ref{series simple-type}) -- is necessarily an \emph{integer} (a fact that will be elucidated shortly). Consequently, the exponential factor in  (\ref{series simple-type}) will imply that $2(S, \lambda) + S^2/2  + f(\chi + \sigma) = 0$ and  $(\tD, 2 \lambda) = i \pi \mathbb Z$; the former condition will hold as long as the operator $I'_2(S)$  in (\ref{I's zero-modes}) is  normalized correctly\footnote{For any $\bar S \in H_2(X\backslash \Sigma, \mathbb R)$, let $r$ be the correct normalization factor (of the classical zero-mode wavefunction) of the operator $I'_2(\bar S)$; from (\ref{I's zero-modes}), it will mean that $I'_2(S)$ -- where $S = r \cdot \bar S$ -- is a correctly normalized version of $I'_2(\bar S)$; the condition $2(S, \lambda) + S^2/2  + f(\chi + \sigma) = 0$ can then be written as $a r^2 + br  + c = 0$, where $a = {\bar S}^2$, $b = 4(\bar S, \lambda)$ and $c = 2f(\chi + \sigma)$ are real constants for any particular choice of $X$ and $\bar S$ such that a solution of $r$ can always be found -- in other words, the condition $2(S, \lambda) + S^2/2  + f(\chi + \sigma) = 0$ will hold if the operator $I'_2(S)$ is normalized  correctly, and vice-versa.}  (a physical requirement that was implicit in our discussions hitherto), and the latter condition just reflects the fact that one is free to appeal to a ``ramification''-preserving, twisted $U(1)$-valued gauge transformation (\ref{twisted gauge tx}) which shifts $\alpha$ in a way compatible with (\ref{condition}).\footnote{Since $2\lambda \in H^2(X, \mathbb Z)$, we can write $\alpha (\Sigma, 2\lambda) = \alpha \sum_i (\Sigma, U_i)$, where $U_i \in H_2(X, \mathbb Z)$ and $(\Sigma, U_i) \in \mathbb Z$. Via (\ref{twisted gauge tx}), one can shift $\alpha$ to satisfy (\ref{condition}); in particular, one can regard $\alpha (\Sigma,  2 \lambda)$ as an even integer so that $(\tD, 2 \lambda) = i \pi \alpha (\Sigma,  2 \lambda) /2 =  i \pi \mathbb Z$ under such a gauge symmetry.} Altogether, it will mean that 
\be
SW(\frak s) =  \langle 1 \rangle_{k'}
\label{proof of Taubes}
\ee
up to a sign, where $\frak s = -i\lambda$ is the corresponding \emph{ordinary} $\spin$-structure.

\bigskip\noindent{\it The Seiberg-Witten Invariants Are The Gromov-Taubes Invariants}

\def\pf{{ \langle 1 \rangle_{k'}}}

What we would like to do now is to determine $\pf$ of (\ref{proof of Taubes}) explicitly. To this end, first note that the parameter $\eta$ in $S_E$ of (\ref{S}) must be set to zero since we are considering nontrivially-embedded surface operators $\Sigma$; then, via a chiral rotation of the massless fermions in the theory which inconsequentially shifts $\Theta$ in  (\ref{S}) to a convenient value, we can write 
\begin{eqnarray}
S_E & = & {1 \over e^2}\int_X d^4 x \, \sqrt{\bar g} \, {\rm Tr} \left[ {1\over 4} F'_{\mu \nu}F^{\mu \nu'} + \half \phi \nabla'_\mu \nabla^{\mu'} \phi^{\dagger} - i \zeta \nabla'_{\mu} \psi^{\mu} - i   \chi^{\mu \nu}_+ \nabla'_{\mu} \psi_{\nu} \right. \nonumber \\
&&  \left. \hspace{0 cm}  -  {i \over 8} \phi [\chi^+_{\mu \nu}, \chi^{\mu \nu}_+] - {i \over 2} \phi^{\dagger} [\psi_\mu, \psi^\mu] - {i \over 2} \phi [\zeta, \zeta] - {1 \over 8} [\phi, \phi^{\dagger}]^2 \right].
\label{Se exp}
\end{eqnarray}

Second, let $\Phi = (A', \phi, \phi^{\dagger})$ and $\Psi = (\zeta, \chi^+, \psi)$ represent the bosonic and fermionic fields of the theory, respectively. Since the semiclassical approximation is exact, one can expand $S_E$ to lowest order in $e$ whence terms beyond quadratic order in the non-zero modes $\tilde \Phi$ and $\tilde \Psi$ need \emph{not} be considered; as such, because there are, for $N_{\psi} = 0$, no zero-modes of $\psi$ and $(\phi, \phi^{\dagger}, \zeta, \chi^+)$ (as explained in $\S$2.3), one can ignore the non-kinetic terms in (\ref{Se exp}) which are beyond quadratic order in $\tilde \Phi$ and $\tilde \Psi$; thus, we can write
\be
S_E = \int_X d^4 x \,  \sqrt {\bar g} \, \left(\tilde \Phi \Delta_B \tilde \Phi +  \tilde \Psi \Delta_F \tilde \Psi \right), 
\ee
where $\Delta_B$ and $\Delta_F$ are certain second and first order elliptic operators, respectively. Hence, the Gaussian integrals over $\tilde \Phi$ and $\tilde \Psi$ will be given by 
\be
{{\rm det} (\Delta_F) \over {\sqrt{{\rm det} (\Delta_B)}}}.
\label{gaussian}
\ee
Note at this point that due to supersymmetry, there is a pairing of the excitations of the fields $\Phi$ and $\Psi$ at every non-zero energy level. Moreover, it is a fact that ${\rm det} (\Delta_F) = {\rm det} (\Delta^{1/2}_B)$ (after one fixes a sign ambiguity by specifying an orientation of the underlying moduli space $\CM'$ of ``ramified'' instantons). Consequently, we have
\be
\Delta_F {\tilde\Psi}_n = \xi_n {\tilde \Psi}_n
\ee
and
\be
\Delta_B {\tilde \Phi}_n = \xi^2_n {\tilde \Phi}_n,
\ee
where the subscript ``$n$'' refers to the $n$-th energy level with corresponding  real eigenvalue $\xi_n \neq 0$. Therefore, one can compute (\ref{gaussian}) to be
\be
\prod_{n} {\xi_n \over \sqrt{|\xi_n|^2}} = \pm 1 = {\rm sign} ({\rm det} \, \Delta_F). 
\label{sign Df}
\ee

Third, recall from our discussions in $\S$2.3 that the non-vanishing contributions (\ref{sign Df}) to $\pf$ localize around ``ramified'' instantons which satisfy (\ref{ram instantons}). Moreover, according to our discussions in $\S$2.3, since there are no $\zeta$ and $\chi^+$ zero-modes, we have ${\rm ind}(\Delta_F) = {\rm dim}(\CM')$ and ${\rm ker}( \Delta_F) = T\CM'$. 

Let us now send the effective value of $\alpha$ to $+1$; i.e., we now have an ``ordinary'' surface operator along $\Sigma$ whence the relation (\ref{proof of Taubes}) is exact.\footnote{When $\alpha =  +1$, one can (as was done earlier in computing (\ref{OS proof compare})) set $\tD = i \pi \alpha \Sigma /2$ to zero in the sign $e^{(\tD, 2 \lambda)}$ of (\ref{proof of Taubes}) via the gauge transformation (\ref{twisted gauge tx}) where $u=-1$.} Then, from the above three points, the fact that $N_{\psi} = {\rm dim} (\CM') = 0$, and the relations (\ref{sign Df}), (\ref{ram instantons}) and (\ref{PD sigma}), one can  conclude that in this case, 
\be
{\langle 1 \rangle}_{k'} = \sum_x \, {\rm sign} ({\rm det} \, {\cal D}),
\label{AJ}
\ee
where $\cal D$ is a certain first-order elliptic operator whose kernel is the tangent space to the space $H$ of solutions to the relation
\be
c_1({\cal E}) = \delta_{\Sigma}.
\label{H}
\ee
 In (\ref{AJ}), the $x$'s are just the points which span the space $H$ of dimension \emph{zero};  in (\ref{H}), $\cal E$ is some nontrivial complex line bundle with a self-dual $\frak {u}(1)$-valued connection $A_{\cal E}$ and curvature $c_1(\cal E)$: recall from our discussion in $\S$2.2 that a choice of an extension of $E$ over $\Sigma$ results in $\alpha$ being $\frak t$-valued, and since the maximal torus $\mathbb T$ of $SU(2)$ is actually $U(1)$, $\alpha$ and therefore $F_+$ (i.e., $2 \pi c_1(\cal E)$) are actually ${\frak u}(1)$-valued  in (\ref{ram instantons}).  Such a complex line bundle $\cal E$ -- where $c_1({\cal E}) \cdot c_1({\cal E}) > 0$ -- can always be found, as $b^+_2 > 1$. Since ${\rm sign} ({\rm det} \, {\cal D}) = \pm 1$, (\ref{AJ}) will imply that $\pf$ is an integer, consistent with (\ref{proof of Taubes}). 

Notice that the relation (\ref{H}) means that one can interpret $H$ as the space of pseudo-holomorphic curves in $X$ whose Poincar\'e dual is $c_1(\cal E)$; hence, with the above description of the kernel of the first-order elliptic operator $\cal D$, one can further conclude that 
\be
 \sum_x \, {\rm sign} ({\rm det} \, {\cal D}) = {\rm Gr}(c_1(\cal E)),
 \label{Gr}
\ee
where ${\rm Gr}(c_1(\cal E))$ is the Gromov-Taubes invariant defined in~\cite{CT ps} for a connected, non-multiply-covered, pseudo-holomorphic curve in $X$ whose fundamental class is Poincar\'e dual to $c_1(\cal E)$.  Since ${\rm Gr}(c_1(\cal E))$ depends only on the homology class of $\Sigma$, it will be invariant under smooth deformations of the metric and complex structure on $\Sigma$; i.e., ${\rm Gr}(c_1(\cal E))$ is also a two-dimensional topological invariant of $\Sigma$. 

On the other hand, because (\ref{proof of Taubes}) is valid for $X$ of SW simple-type, i.e., $\lambda^2 - (2 \chi + 3 \sigma) /4 = d_{L^2_d}  = - c_1(L_d^2)[\Sigma] +  \Sigma \cap \Sigma = 0$, we find that (\ref{SW = SW}), (\ref{exp SW}) and ${\frak s'} = \frak s - \delta_\Sigma$ will imply that the LHS of (\ref{proof of Taubes}) is $SW(\frak s - \delta_\Sigma)$. In fact, since the non-vanishing contributions to the RHS of (\ref{proof of Taubes}) localize around supersymmetric field configurations which obey (\ref{H}), the LHS of  (\ref{proof of Taubes}) can actually be written as $SW(\frak s - c_1(\cal E))$;  as a result,  by (\ref{Gr}), (\ref{AJ}) and (\ref{proof of Taubes}), we have
\be
SW(\frak s - c_1({\cal E})) =   {\rm Gr}(c_1(\cal E)). 
\label{taubes}
\ee 

In any case, because $\Sigma$ is such that $\Sigma \cap \Sigma > 0$, it must satisfy (cf.~(\ref{result now}))
\be
2 -2g + \Sigma \cap \Sigma \leq - c_1(L^2_d) [\Sigma],
\label{inequality new}
\ee
in addition to (\ref{adj Taubes});  consequently, we necessarily have $\frak s = {1 \over 2} c_1(L^2_d) = {1 \over 2} c_1(K)$. By noting that as $b^+_2 > 1$, the ordinary SW invariants satisfy $SW( \bar {\frak s}) = \pm SW( - \bar {\frak s})$ for any ordinary $\spin$-structure $\bar {\frak s}$~\cite{monopoles},  we can also write (\ref{taubes}) as
\be
SW (\hat {\frak s}) = \pm  {\rm Gr}(c_1(\cal E)),
\label{prt}
\ee
where 
\be
\hat {\frak s} = {1 \over 2} c_1(\mathscr L),
\label{prt 1}
\ee
and 
\be
{\mathscr L} = K^{-1} \otimes {\cal E}^2. 
\label{prt 2}
\ee
Moreover, since $c_1(L^2_d) = c_1(K)$,  the condition $d_{L^2_d} =0$  can also be expressed as
\be
- c_1(K) \cdot c_1({\cal E}) +   c_1({\cal E}) \cdot  c_1({\cal E}) =0.
\label{d condition}
\ee
Because the LHS of (\ref{d condition}) is the dimension of $H$ as defined mathematically in~\cite{CT ps}, we see that (\ref{d condition}) is indeed consistent with the fact that  $H$ is zero-dimensional as implied by $N_{\psi} = 0$.  Moreover, (\ref{d condition}) and (\ref{adj Taubes}) together imply that the genus of the pseudo-holomorphic curve represented by $c_1(\cal E)$ will be given by  
\be
g = 1 + c_1({\cal E}) \cdot c_1(\cal E). 
\label{gps}
\ee
  
Finally, note that (\ref{prt})-(\ref{gps}) are \emph{precisely} Theorem 4.1 and Propositions 4.2-4.3 of~\cite{MRL} which summarizes the results collected in~\cite{CT 1}!  This completes our physical derivation of Taubes' groundbreaking result that the ordinary SW invariants  are  (up to a sign) equal to the Gromov-Taubes invariants  on any compact, oriented symplectic four-manifold with $b^+_2 > 1$.  

\newsection{Mathematical Implications Of The Underlying Physics}

Now that we have physically re-derived the above mathematically established theorems by Ozsv\'ath-Szab\'o and Taubes, one might wonder if the physics can, in turn, offer any new and interesting mathematical insights. Indeed it can, as we shall now elucidate.

\newsubsection{The Gromov-Taubes And ``Ramified'' Seiberg-Witten Invariants}

\def\fD{{\frak D}}

Assume that $X$ is a compact, oriented, symplectic four-manifold which contains at least one trivially-embedded curve that is connected; assume also that $b_1(X)=0$ and  $b^+_2(X) > 1$. Then, from (\ref{prt}), and (9.9) of~\cite{u-plane SO}, we find that 
\be
SW (\hat {\frak s}_{\rm r} + {\alpha_{\rm r}} \delta_{\fD}) = \pm  {\rm Gr}(c_1(\cal E)).
\label{GW=SW}
\ee
Here, $SW (\hat {\frak s}_{\rm r} + {\alpha_{\rm r}} \delta_{\fD})$ is a ``ramified'' SW invariant; in addition, the gauge field underlying $\hat {\frak s}_{\rm r}$ picks up a \emph{nontrivial }holonomy -- parameterized by a non-integer $\alpha_{\rm r}$ -- as one traverses a closed loop linking a connected curve $\fD$ that is \emph{ trivially-embedded } in $X$. In other words, the Gromov-Taubes invariants -- which count the connected, non-multiply-covered, pseudo-holomorphic curves with positive self-intersection and fundamental class Poincare dual to $c_1(\cal E)$ -- are (up to a sign) equal to the ``ramified'' Seiberg-Witten invariants of $X$!

\bigskip\noindent{\it A Rigorous Mathematical Proof?} 

Let us now attempt to explain why (\ref{GW=SW}) ought to be amenable to a rigorous mathematical proof. First, note that the presence of a bona-fide ``ramification'' along $\fD$ implies that the LHS of (\ref{GW=SW}) counts (with signs)  the number of solutions to the ``ramified'' SW equations which are defined (in the mathematical convention) by
\be
F_+ = (\overline M M)_+  -   \mu  
\label{Taubes SW 1}
\ee
and 
\be
\dirac M =0.
\label{Taubes SW 2}
\ee
Here, $F$ is an imaginary-valued, curvature two-form given by $F = - i \pi c_1({\mathscr L}_F)$, where ${\mathscr L}_F = \mathscr L \otimes {\mathscr L}^{2\alpha_{\rm r}}_\fD$; the complex line bundle $\mathscr L_\fD$ is such that $c_1({\mathscr L}_\fD)$ is the Poincar\'e dual $[\fD]$ of $\fD$;  $\mu =  i \epsilon \delta^+_\fD$,  where $\epsilon$  is a positive real constant, is a fixed, imaginary-valued, self-dual two-form on $X$ that cannot be set to zero; and $M$ is a section of the complex vector bundle $S_+ \otimes {\mathscr L}^{1/2}$, where $\mathscr L$ is as given in (\ref{prt 2}). 

Second, notice that (\ref{Taubes SW 1}) is the same as
 \be
 F_+ = (\overline M' M')_+  -   \mu, 
 \label{alt 1}
 \ee
where  ${\overline M}' = \overline M e^{i \alpha_{\rm r} \theta}$ and $M' = e^{-i \alpha_{\rm r} \theta} M$ are gauge-transformed versions of $\overline M$ and $M$, respectively. Notice also that (\ref{Taubes SW 2}) is the same as 
\be
\dirac' M' = 0,
\label{alt 2}
\ee
where (assuming a small but non-zero $\alpha_r$)  $\dirac' M' = e^{-i \alpha_{\rm r} \theta} \dirac M$ such that $M'$ is a section of the complex vector bundle $S_+ \otimes {\mathscr L}_F^{1/2}$. Altogether, this means that one can interpret the ``ramified'' SW equations of (\ref{Taubes SW 1})-(\ref{Taubes SW 2}) as the ordinary, \emph{perturbed} SW equations of (\ref{alt 1})-(\ref{alt 2}) with perturbation two-form $\mu$ and  $\spin$-structure ${\frak s}_{\rm o} = {1\over 2} c_1({\mathscr L}_F)$.  Moreover, via (\ref{prt 2}), one can also write ${\mathscr L}_F = K^{-1} \otimes {\cal E}^2_F$, where ${\cal E}_F = {\cal E} \otimes {\mathscr L}^{\alpha_{\rm r}}_\fD$. 

Third, note that since $\fD \cap \fD =0$,  it will mean that $\fD \neq \Sigma$, where $\Sigma \subset X$ is the connected, psuedo-holomorphic curve introduced at the start of $\S$4 with \emph{positive} self-intersection. In particular, $[\fD]$ and $[\Sigma]$  are necessarily distinct. Consequently, since $c_1({\cal E}) = [\Sigma]$, it must be that $ {\mathscr L}_\fD \neq {\cal E}$. 

Fourth, notice that for a fixed $\fD$ and $\alpha_{\rm r}$,  the map $\Sigma \to [\Sigma]$ is potentially many-to-one while the map $\fD \to [\fD]$ is necessarily one-to-one. As such, there is a one-to-one correspondence between (pseudo-holomorphic) curves in $X$ whose Poincar\'e duals are $c_1({\cal E}_F) =  [\Sigma] + \alpha_{\rm r} [\fD]$ and $c_1({\cal E})  = [\Sigma]$.

Last but not least, note that $\mu  =  i \epsilon  \delta^+_\fD$ in (\ref{alt 1}) is singular: this is because $\delta_\fD$ is actually a delta two-form. Hence, according to Taubes' analysis in~\cite{CT 1}, each solution to (\ref{alt 1})-(\ref{alt 2}) ought to determine a pseudo-holomorphic curve in $X$ that is Poincar\'e dual to $c_1({\cal E}_F)$. Consequently, the above-observed one-to-one correspondence  between pseudo-holomorphic curves in $X$ whose Poincar\'e duals are $c_1({\cal E}_F)$ and $c_1({\cal E})$, will imply that each solution to (\ref{alt 1})-(\ref{alt 2}) ought to correspond to a pseudo-holomorphic curve in $X$ that is Poincar\'e dual to $c_1({\cal E})$.  This conclusion is indeed consistent with (\ref{GW=SW}).

\newsubsection{Certain Identities Among The Gromov-Taubes Invariants}

Assume that $X$ is a compact, oriented, symplectic four-manifold with $b_1=0$ and $b^+_2  > 1$. Then, from (\ref{proof of Taubes}), (\ref{AJ}) and (\ref{Gr}), we have 
\be
SW({\frak s}) =  {\rm Gr} (c_1({\cal E})),
\label{GT 1}
\ee
where as explained in $\S$4, we necessarily have 
\be
{\frak s} = {1\over 2} c_1(K^{-1} \otimes K^2)
\label{GT 2}
\ee 
if $c_1(\cal E)$ is the Poincar\'e dual of  the pseudo-holomorphic curve $\Sigma$. 

Via (\ref{GT 1}), (\ref{GT 2}) and (\ref{prt})-(\ref{d condition}), we find that 
\be
{\rm Gr} (c_1(K)) = \pm {\rm Gr} (c_1(\cal E)).  
\label{GT 3}
\ee
Note that  the dimension of the space of pseudo-holomorphic curves associated with the LHS  of (\ref{GT 3})  is given by  (\ref{d condition}), albeit with $\cal E$ replaced by $K$; in particular, it is zero, just like the dimension of the space of pseudo-holomorphic curves associated with the RHS of (\ref{GT 3}). In this sense, (\ref{GT 3}) can be viewed as a consistent relation. But can we say more? Most certainly. 

Firstly, it is clear that (\ref{GT 3}) implies that there exists pseudo-holomorphic curves in $X$ which are Poincar\'e dual to $c_1(K)$. Since pseudo-holomorphic curves (in a symplectic four-manifold) are automatically symplectic~\cite{CT 1}, it will mean that the Poincar\'e dual of $c_1(K)$ can be represented by a fundamental class of an embedded symplectic curve in $X$; this conclusion is just Theorem 0.2 in article 1 of~\cite{CT 1}.  Moreover, this conclusion also implies via (\ref{gps}) that  if there are no embedded spheres in $X$ with self-intersection $-1$, then $c_1(K) \cdot c_1(K) \geq 0$; this observation agrees with Proposition 4.2 of~\cite{MRL}. 

Secondly, (\ref{GT 3}) also implies that $c_1(\cal E)$ is represented by at least one pseudo-holomorphic curve in $X$ -- a fact that is well-established in the mathematical literature~\cite{wild}.

In any event, the above mathematical assertions depend squarely on the non-vanishing of  ${\rm Gr} (c_1(\cal E))$; thus, one can understand them to be a consequence of $R$-invariance: $R$-invariance of the topological partition function $\langle 1 \rangle_{k'}$ of the $k'$-instanton sector asserts that it will not vanish, and from (\ref{AJ}) and (\ref{Gr}), we see that ${\rm Gr} (c_1({\cal E}))$ will not vanish either. 

Notice also that since ${\rm Gr}(0) =1$ by definition~\cite{CT 1}, we have $SW(\frak s) = 1$ from (\ref{taubes}). Consequently, (\ref{GT 1}) will mean that
\be
{\rm Gr} (c_1({\cal E})) = + 1. 
\label{Gr =1}
\ee  
In other words, the number of points in the zero-dimensional space $H$ of pseudo-holomorphic curves $\Sigma \subset X$ which are positively-oriented is greater than the number which are negatively-oriented by \emph{one}.  In fact, (\ref{Gr =1}) is consistent with the relation ${\rm Gr}_0(A) = \pm 1$ proved in Proposition 3.18 of~\cite{dusa}, while (\ref{GT 3}) -- in light of (\ref{Gr =1}) -- is consistent with the relation $\vert {\rm Gr}(K) \vert = 1$ proved in Theorem 3.10 of~\cite{dusa}. 

Last but not least, since $SW(\bar {\frak s}) = \pm SW(-\bar {\frak s})$ for any ordinary $\spin$-structure $\bar {\frak s}$, from (\ref{prt})-(\ref{prt 2}) and (\ref{GT 3}), we find that
\be
{\rm Gr} (c_1(K)) = \pm {\rm Gr} (c_1(K) - c_1({\cal E})).
\label{GT 4}
\ee
Note that (\ref{GT 4}) is distinct from the widely-known result of Serre-Taubes duality for pseudo-holomorphic curves in $X$~\cite{wild} (which one can nevertheless obtain from (\ref{GT 4}) by  making the substitution (\ref{GT 3})).   
 
In summary, for \emph{connected},\emph{ non-multiply-covered}, pseudo-holomorphic curves in $X$ that have \emph{positive} self-intersection and fundamental class Poincar\'e dual to $c_1(\cal E)$,   the underlying physics suggests that the relations (\ref{GT 3}), (\ref{Gr =1}) and (\ref{GT 4}) ought to hold in addition to those which have already been established in the mathematical literature.

\newsubsection{Affirming A Knot Homology Conjecture By Kronheimer And Mrowka}

\def\cA{{{\cal A}' / {\cal G}'}}

\def\mA{{\mathscr A'}}
	
\def\cH{{H}}

\def\bcq{{\overline {\cal Q}}}

Assume that \emph{general} $X = M \times {{\bf S}^1}$, where $M$ is a compact, oriented three-manifold, and $b_1(M) = b^+_2 (X) > 1$. Recall that the effective Lagrangian of the topological $\CN =2$ pure $SU(2)$ gauge theory with an arbitrarily-embedded surface operator $\Sigma$, is just the Lagrangian of the ordinary Donaldson-Witten theory with gauge field $A'$ and field strength $F' = F - 2\pi i \alpha \delta_\Sigma$. As such, one can conclude from the analysis in~\cite{21} that up to $\cq$-exact terms which are thus irrelevant, $S_E$ in (\ref{Se exp}) is the action\footnote{Recall from $\S$2.2 that unless the surface operator is nontrivially-embedded, there is\emph{ no} restriction on the effective values that its $\eta$-parameter can take in order to preserve modular invariance in the corresponding low-energy SW theory. As such, let us for simplicity, take $\eta$ to be zero in our following analysis; then, $S_E$ in (\ref{Se exp}) will be the relevant action regardless of the embedding of the surface operator in $X$.} for a supersymmetric quantum mechanical sigma model with target manifold $\cA$ -- the space of all gauge-inequivalent classes of ``ramified'' $SU(2)$-connections $\mA$ on $M$,  and potential $h = \half \int_{M} {\rm Tr} \, (\mA \wedge d\mA + {2\over 3} \mA \wedge \mA \wedge \mA)$ -- the Chern-Simons functional of $\mA$. Specifically, $\mA$ can be regarded a gauge connection of an $SU(2)$-bundle over $M \backslash \Sigma_M$ -- where $\Sigma_M \subset \Sigma$ is the component of $\Sigma$ embedded in $M$ -- whose holonomy around a small circle linking $\Sigma_M$  in $M$ is ${\rm exp} (2 \pi i \alpha)$.  Moroever, we now have in the supersymmetry algebra a Hamiltonian operator $\cH$ which generates translations in the ``time'' direction along ${{\bf S}^1}$, and a second nilpotent supercharge $\bcq$. In particular,  they obey $[\cH, \cq] = [\cH, \bcq] = 0$ and  $\{ \cq, \bcq\} = 2 \cH$; consequently, one can easily show that the ground states of the theory are supersymmetric, i.e., they must be annihilated by both $\cq$ and $\bcq$, and that they are in the $\cq$-cohomology. (See chapter 10 of~\cite{MS} for an excellent review of this and other assertions to be made momentarily.)

What we would like to do now is to compute the partition function $\pf$ of the theory via the supersymmetric quantum mechanical sigma model on $\cA$. Since the presence of ${{\bf S}^1}$ in $X$ enforces a periodic boundary condition on the (fermi) fields of the theory, the path-integral of the sigma model without operator insertions, i.e., $\pf$, will be given by the Witten index ${\rm Tr} (-1)^F$, where $F$ is the fermion number.  In turn, ${\rm Tr} (-1)^F$ is given by the Euler characteristic of the $\cq$-complex generated by the $\cq$-cohomology groups, i.e., the supersymmetric ground states.  

As first pointed out by Atiyah in~\cite{Weyl}, in the case that one has an ordinary $SU(2)$ connection $\mathscr A$ on (a homology three-sphere) $Y$,  the ground states of the corresponding Hamiltonian are, purely formally, the instanton Floer homology groups $HF_\ast(Y)$ of $Y$ defined by Floer~\cite{Floer}. Analogously, as first suggested in~\cite{sergei}, one can formally identify the ground states of $\cH$ as the ``ramified'' instanton Floer homology groups  $HF_\ast(M; \Sigma_M; \alpha)$ of $M$; $\pf$ will then be given by the Euler characteristic of the ``ramified'' instanton Floer homology of $M$.  In other words, we have
\be
\pf = \chi(HF_\ast(M; \Sigma_M; \alpha)).
\label{pf=chi}
\ee   
Moreover, it was also verified in~\cite{u-plane SO} that the $R$-symmetry under which $\cq$ has charge 1 is only conserved mod 8; as a result, the ``ramified'' instanton Floer complex, like its ordinary counterpart, has a mod 8 grading under this $R$-symmetry. Nevertheless, unlike its ordinary counterpart, its relative grading is defined mod 4 instead of mod 8.\footnote{The relative grading, as defined mathematically for the ordinary instanton Floer complex, depends on the index which computes the dimension of the moduli space $\cal M$  of $SU(2)$-instantons; in particular, since ${\rm dim}({\cal M}) = 8k - {3 \over 2} (\chi + \sigma)$, where $k$ takes different integer values in different topological sectors, the relative grading is defined mod 8. In this sense, since ${\rm dim}({\cal M}') = 4(2k + l) - {3 \over 2} (\chi + \sigma) - 2(g-1)$, where $k$ and $l$ take different integer values in different topological sectors, the relative grading of the ``ramified'' instanton Floer complex will be defined mod 4.}

That (\ref{pf=chi}) is a consistent relation can be seen as follows. First, note that $\chi(M \times {{\bf S}^1}) = \chi(M)\chi({{\bf S}^1}) =0$; similarly, as $b^+_2 (X) = b^-_2 (X)$, we have $\sigma(M \times {{\bf S}^1}) = 0$; as such, from (\ref{k'}), it will mean that $\pf$ is the partition function of the topologically trivial sector where $k' =0$.  Second, note that for $k' =0$, the dimension of the moduli space of flat ``ramified'' $SU(2)$-connections on $X$ is given by $- 3 \chi (M \times {{\bf S}^1}) = 0$; in other words, there are a discrete number of flat solutions of $A'$; consequently, as $X = M \times {{\bf S}^1}$ is a trivial product of two spaces, each such flat solution of $A'$ on $X$ will correspond to a flat solution of $\mA$ on $M$; hence, the dimension ${\rm dim}(\CM'_f)$ of the moduli space $\CM'_f$ of flat ``ramified'' $SU(2)$-connections $\mathscr A'_{f}$ on $M$, is zero. Third, note that it is well-established that the number of supersymmetric ground states of the sigma model is invariant under rescalings of the potential $h$; therefore, one can rescale $h \to \gamma h$, where $\gamma \gg 1$, and the Witten index ${\rm Tr}(-1)^F$ -- which counts the difference in the number of bosonic and fermionic ground states -- will not change; this means that one can compute $\chi(HF_\ast(M; \Sigma_M; \alpha))$ after such a rescaling of $h$, and still get the correct result. Last but not least, note that when $\gamma \gg 1$,  the contributions to $\chi(HF_\ast(M; \Sigma_M; \alpha))$ will localize onto the critical point set of $h$, i.e., $\CM'_f$. Thus, since ${\rm dim}(\CM'_f) = 0$, i.e., $\CM'_f$ consists of zero-dimensional points only, we have
\be
\chi(HF_\ast(M; \Sigma_M; \alpha)) = \sum_{x} {\rm sign} (h'' (x)) =  \sum_{x} \pm 1, 
\label{points}
\ee
where $h''(x)$ is the Hessian of $h$ at the point $x \in \CM'_f$.\footnote{Since $b_1(M) \neq 0$, one might encounter a situation whereby some of the points in $\CM'_f$ are degenerate. Nevertheless, for an appropriate nontrivial restriction of the $SU(2)$-bundle to $M$, one can -- without altering the Witten index ${\rm Tr}(-1)^F$ and therefore, $\chi(HF_\ast(M; \Sigma_M; \alpha))$ --  perturb $h$  so that its critical point set will consist of a finite number of isolated, non-degenerate and irreducible points which we can then interpret as the $x$'s in (\ref{points}) (cf. Prop. 3.12 of~\cite{dym}).} In particular, the topological invariant $\chi(HF_\ast(M; \Sigma_M; \alpha))$ is -- like $\pf$ computed using (\ref{sign Df}) -- a sum of signed points; an \emph{integer}. It is in this sense that (\ref{pf=chi}) is deemed to be a consistent relation.  

\bigskip\noindent{\it Implications For A Knot Homology Group From ``Ramified'' Instantons}

In fact, one can say more if $\Sigma_M$ is a knot $K \subset M$. In this case, $\chi(HF_\ast(M; K; \alpha))$ in (\ref{points}) counts (with signs) the number of flat $SU(2)$-connections $\mathscr A'_{f}$ on $M \backslash K$ with holonomy ${\rm exp}(2 \pi i \alpha)$ around a circle linking $K$ in $M$. Also, $\mathscr A'_{f}$ only picks up nontrivial contributions to the holonomy along a path that lies in the plane normal to (the singularity along) $K$, i.e., along the $\theta$-direction; therefore, if $K$ is a nontrivial knot -- i.e., if there are crossings that cannot be undone by any orientation-preserving homeomorphism of $M$ to itself -- the holonomy of ${\mathscr A}'_{f}$ along the longitude of $K$ will always be nontrivial (for some judicious choice of the $\alpha$-parameter of the surface operator). Therefore,  one can also interpret $\chi(HF_\ast(M; K; \alpha))$  as an algebraic count of the number of conjugacy classes of homomorphisms 
\be
\rho: \pi_1(M \backslash K) \to SU(2)
\ee
which satisfy the constraint that $\rho$ maps -- via the holonomy of ${\mathscr A}'_{f}$ -- the longitude of $K$ to a \emph{non-identity} element of $SU(2)$. In turn, this implies that the groups $HF_\ast(M; K; \alpha)$ are in one-to-one correspondence with the conjugacy classes $\rho$ with the stated constraint. Thus, we can identify $HF_\ast(M; K; \alpha)$ with the knot homology groups $LI_\ast(M,  K)$ from ``ramified'' $SU(2)$-instantons defined by Kronheimer and Mrowka in $\S$4.4 of~\cite{dym}; in particular, we have $\chi(HF_\ast(M; K; \alpha)) = \chi(LI_\ast(M,  K)) \neq 0$. 

Note at this point that since the partition function $\pf$ is invariant under deformations of the metric on $X$,  the relation (\ref{pf=chi}) would imply that $\chi(HF_\ast(M; K; \alpha))$ and hence $ \chi(LI_\ast(M,  K))$ are invariant under homeomorphisms of $M$ to itself. Consequently, if $K_0$ is an unknot which thus bounds a (twisted) disk in $M$, one can always -- via a suitable orientation-preserving homeomorphism of $M$ to itself -- deform $K_0$ to a trivial unknot  ${\tilde K}_0$ (i.e., a geometrically round circle)  such that $\chi(LI_\ast(M; K_0)) = \chi(LI_\ast(M; {\tilde K}_0))$. As the holonomy of  $\mathscr A'_{f}$ along the longitude of the trivial unknot $\tilde K_0$ can only be the identity element of $SU(2)$ (according to our explanations in the last paragraph), the set of constrained maps $\rho$ in question will be empty for $\tilde K_0$; i.e.,  $LI_\ast(M, \tilde K_0)$ and therefore $\chi(LI_\ast(M; {K}_0))$ are zero. 

In summary, we find that  $\chi(LI_\ast(M, K))$ is zero only if $K$ is an unknot. Therefore, our above analysis physically affirms a mathematical conjecture proposed by Kronheimer and Mrowka in $\S$4.4 of~\cite{dym}, which asserts that $\chi(LI_\ast(M, K))$ vanishes if the symmetrized Alexander polynomial of the knot $K$ is trivial, i.e., if $K$ is an unknot.

\newsubsection{The Gromov-Taubes Invariant, Instanton Floer Homology, And The Casson-Walker-Lescop Invariant}

\bigskip\noindent{\it The Gromov-Taubes Invariant of $M \times {{\bf S}^1}$ And The Instanton Floer Homology of $M$}

Assume that \emph{symplectic} $X = M \times {{\bf S}^1}$, where $M$ is a compact, oriented three-manifold, and $b_1(M) = b^+_2 (X) > 1$. Now, let us consider the surface operator $\Sigma$ to be a pseudo-holomorphic curve in $X$ whose characteristics are as described at the beginning of $\S$4. Let us also send the effective value of $\alpha$ to  +1. Then, according to our analysis in $\S$4, the LHS of (\ref{pf=chi}) will be given by the Gromov-Taubes invariant ${\rm Gr}(c_1(\cal E))$, where  $c_1(\cal E)$ is Poincar\'e dual to $\Sigma$ with positive self-intersection, and $\cal E$ is a complex line bundle with self-dual connection $A_{\cal E}$.

On the other hand, when $\alpha = +1$, the holonomy ${\rm exp}(2 \pi i \alpha)$ of the $SU(2)$ gauge connection $\mA$ around a small circle which links $\Sigma_M$ in $M$, is trivial; in other words, $\mA$ can, in this case, be regarded as an ordinary $SU(2)$ connection on $M$. In turn, this means that one can, in such a situation, replace $HF_\ast(M; \Sigma_M; \alpha)$ on the RHS of (\ref{pf=chi}) with the \emph{ordinary} instanton Floer homology groups $HF_\ast (M)$ of $M$.  

From the preceding two points, one can therefore conclude that 
\be
{\rm Gr}(c_1({\cal E})) = \chi (HF_\ast (M)).
\label{wow}
\ee
In other words, the Gromov-Taubes invariant which algebraically counts \emph{connected}, \emph{non-multiply-covered}, pseudo-holomorphic curves in $M \times {{\bf S}^1}$ with \emph{positive} self-intersection, is equal to the Euler characteristic of the instanton Floer homology of $M$ with $b_1(M) > 1$! 

One can immediately validate (\ref{wow}) for  $M = {\bf T}^3$, since the relevant mathematical results exist. In this case, $X = {\bf T}^3 \times {\bf S}^1$ is symplectic K\"ahler with $ b^+_2(X) = b_1(M) =3$, and according to~\cite{Mun}, $\chi(HF_\ast({\bf T}^3)) = +1$.\footnote{Note that this mathematical result of~\cite{Mun} is actually valid for $G = SO(3)$. However, $\alpha$ continues to take values in ${\frak u}(1)$ when $G = SO(3)$ instead of $SU(2)$; consequently, when $G = SO(3)$, our computations will still lead us to (\ref{wow}) -- i.e., (\ref{wow}) also holds for $G = SO(3)$. Hence, we can still check against this mathematical result.} What about ${\rm Gr}(c_1(\cal E))$? Well, although $b^+_2(X) > 1$, because $b_1(X) > 0$, one cannot read off from our result in (\ref{Gr =1}) (which is defined for $b_1(X) =0$ and $b^+_2(X) > 1$). However, from the relation ${\rm Gr}_0(A) = \pm 1$ proved in Proposition 3.18 of~\cite{dusa},  and the fact that on any K\"ahler manifold, the almost complex structure $J$ is necessarily integrable and thus, all points in the space of pseudo-holomorphic curves have positive orientation, i.e., all points contribute as $+1$ in the computation of ${\rm Gr}(c_1(\cal E))$~\cite{dusa}, we can conclude that ${\rm Gr}(c_1({\cal E})) = +1$ on $X$, too. Therefore, we have $\chi(HF_\ast({\bf T}^3)) = {\rm Gr}(c_1({\cal E})) = +1$, which certainly agrees with (\ref{wow}). 

In fact, one can validate (\ref{wow}) for \emph{any} $M = \Sigma_g \times {\bf S}^1$, where $\Sigma_g$ is a compact Riemann surface of genus $g > 1$. To this end, first note that $X$ is, in this case, a minimal symplectic manifold with $b^+_2(X) = b_1(M) = 1 + 2g$. Other than the $g=1$ example above, its canonical bundle $K$ is nontrivial; however, since $g >1$, $X$ is an elliptic surface of Kodaira dimension 1 -- i.e., $K^2 = 0$~\cite{friedmann}. Consequently, because $c_1({\cal E})^2  > 0$, Theorem 3.10\,(\textit{iv}) of~\cite{dusa} will imply that ${\rm Gr}(c_1({\cal E})) = 0$. At the same time,  we have $\chi(HF_\ast(M)) = 0$ for $g > 1$~\cite{Mun}. In summary, for the elliptic surfaces $X = \Sigma_g \times {\bf T}^2$ where $g \geq 1$,  (\ref{wow}) is found to be consistent with all known mathematical results. 

Another nontrivial check on the validity of (\ref{wow}) is as follows. Let $\Sigma = K_0 \times {\bf S}^1$, where $K_0 \subset M$ is an unknot whence $\Sigma$ is homeomorphic to a genus one curve in $X$. Recall from our discussion at the beginning of $\S$4 that in our case, the topological invariant ${\rm Gr}(c_1(\cal E))$ does \emph{not} count curves of genus one in $X$; in other words, ${\rm Gr}(c_1({\cal E})) = 0$ for such a $\Sigma$. At the same time, for such a $\Sigma$, we have $ \chi (HF_\ast (M))  =  \chi(HF_\ast(M; K_0; 1)) = 0$ from our discussion in the previous subsection. Again, this observation agrees with (\ref{wow}).

\bigskip\noindent{\it The Gromov-Taubes And The Casson-Walker-Lescop Invariants}

Let us also mention that it was argued in~\cite{moore-marino} that $\chi (HF_\ast (M)) = \lambda_{CWL}(M)$, where $\lambda_{CWL}(M)$ is the Casson-Walker-Lescop invariant of $M$~\cite{lescop}; for example, one has $\chi (HF_\ast ({\bf T}^3)) = \lambda_{CWL}({\bf T}^3) = +1$.  Hence, (\ref{wow}) will imply that for symplectic $X = M \times {\bf S}^1$,
\be
{\rm Gr}(c_1({\cal E})) = \lambda_{CWL}(M).
\label{cwl}
\ee
Consequently,  since $ \lambda_{CWL}(M) = 0$ if $b_1(M) > 3$, it will mean that
\be
{\rm Gr}(c_1({\cal E})) = 0, \quad {\rm if} \quad b^+_2(X) > 3.  
\label{van}
\ee
Notice that (\ref{cwl}) and (\ref{van}) indeed agree with our analysis of $X = \Sigma_g \times {\bf T}^2$ above. 

\def\cM{{\cal M}}

\newsubsection{The Monopole Floer Homology And  Seiberg-Witten Invariants Of Three-Manifolds}

\medskip\noindent{\it A Relation Between The Instanton And Monopole Floer Homologies Of $M$}

Again, let us consider $X = M \times {\bf S}^1$ to be \emph{symplectic}, where $M$ is a compact, oriented, three-manifold with $b_1(M) = b^+_2(X)  > 1$.  In this case, the relation (\ref{wow}) also leads to an important implication for a Seiberg-Witten or monopole Floer homology group $HM_\ast(Y, {\frak s}_Y)$  of a general three-manifold $Y$ with $\spin$-structure ${\frak s}_Y$  described by Kronheimer in~\cite{PB}.\footnote{Other variants of this monopole homology group were subsequently defined and constructed by Kronheimer-Mrowka in~\cite{3-manifolds}; they were also studied in detail by Kutluhan-Taubes in~\cite{SWF} for when $Y = M$, $b_1(M) > 1$ and $X = M \times {\bf S}^1$ is symplectic  -- in other words, our case at hand.} This can be understood as follows.  

First, let us denote $SW(X, {\frak s})$ as the Seiberg-Witten invariant of $X$ determined by a $\spin$-structure $\frak s$ on $X$; 
 let us also denote $SW(M, {\frak s}_M)$ as the Seiberg-Witten invariant of $M$ (which ``counts'' the number of solutions of the three-dimensional Seiberg-Witten equations on $M$ obtained by dimensional reduction along ${\bf S}^1$ of the original four-dimensional Seiberg-Witten equations on $X$)  determined by a $\spin$-structure ${\frak s}_M$ on $M$;  then, one can prove that $SW(X, \pi^{-1}({\frak s}_M)) = SW(M, {\frak s}_M)$, where $\pi:  M \times {\bf S}^1  \to M$~\cite{PB}. Second, note that $\chi (HM_\ast(M, {\frak s}_M)) = SW(M, {\frak s}_M)$~\cite{marcolli}. Third, recall that  as explained in footnote~16, (\ref{wow}) is also valid for when the gauge group underlying $HF_\ast(M)$ is $SO(3)$. Altogether, this means that (\ref{wow}), in light of (\ref{prt}), will imply that up to a sign, we have an equivalence
\be
\chi(HM_\ast(M, \pi(\hat {\frak s}))) = \chi (HF^{w}_\ast(M))
\label{chi=chi}
\ee
between the Euler characteristics of the monopole and instanton Floer homologies of $M$! Here, $w = w_2(E_M)$ is the second Stiefel-Whitney class of the $SO(3)$ gauge bundle $E_M$ over $M$, and according to the Main Theorem of~\cite{SWF} and $\S$40.1 of~\cite{3-manifolds}, the first Chern class of the projection $\pi(\hat {\frak s})$ to $M$ of the $\spin$-structure $\hat {\frak s}$ on $X$ is necessarily non-torsion. (Recall that $c_1(\hat {\frak s}) = c_1({\cal E}) - {1\over 2} c_1(K)$, where $K$ is the canonical line bundle of $X$, and $c_1({\cal E})$ is the Poincar\'e-dual of the fundamental class of the \emph{connected}, \emph{non-multiply-covered} pseudo-holomorphic curve $\Sigma$ with \emph{positive} self-intersection.)

From (\ref{chi=chi}), it is also clear that if the monopole Floer homology $HM_\ast(M, \hat {\frak s}_M)$ is nontrivial for some $\spin$-structure $\hat {\frak s}_M$ whose first Chern class is non-torsion, then the instanton Floer homology $HF^{w}_\ast(M)$ is nontrivial too.  In the case that $M = \Sigma_{g} \times {\bf S}^1$ with $g \geq 1$,  this result is just a generalization of Conjecture 6.3 in~\cite{PB} proposed by Kronheimer for $g=0$.   

In fact, for $M = \Sigma_g \times {\bf S}^1$ with $g \geq 1$, it was proved in~\cite{munoz 1} that $HF^w_{\ast}(M)$ is isomorphic to the quantum cohomology $QH^\ast(\cM_{\Sigma_g})$ of the moduli space $\cM_{\Sigma_g}$ of flat $SO(3)$-connections on $\Sigma_g$ with nontrivial second Stiefel-Whitney class $w$; it was also proved in~\cite{munoz 2} that  $HM_\ast(M, \pi(\hat {\frak s}))$ is isomorphic to the quantum cohomology $QH^\ast(s^r(\Sigma_g))$ of the $r$-symmetric product $s^r(\Sigma_g)$ of the Riemann surface $\Sigma_g$, where the integer $r$ is related to the choice of the $\spin$-structure $\pi(\hat {\frak s})$; since it is shown in~\cite{thaddeus} that the space $\cM_{\Sigma_g}$ can be smoothly linked to the space $s^r(\Sigma_g)$,  one ought to be able to identify $HF^w_{\ast}(M)$ with $HM_\ast(M, \pi(\hat {\frak s}))$ such that (\ref{chi=chi})  will hold. In short, for  $M = \Sigma_{g} \times {\bf S}^1$ with $g \geq 1$, (\ref{chi=chi}) is found to be consistent with expectations from existing mathematical results.

\bigskip\noindent{\it Topology Of The Moduli Space Of Flat $SU(2)$-Connections On $M$}

Another implication of (\ref{chi=chi}) can be understood as follows. First, note that  $SW(X_4, \hat {\frak s})$ vanishes identically if the four-manifold $X_4$ has positive scalar curvature and $b^+_2(X_4) > 1$~\cite{monopoles}; in our case, this will mean that $SW(X, \hat {\frak s}) = SW(M, \pi (\hat {\frak s})) = \chi(HM_\ast(M, \pi(\hat {\frak s}))) = 0$ identically if $M$ has positive scalar curvature. Second, note that $ \chi(HF_\ast(M; \Sigma_M; \alpha)) = \chi(HF^{w=0}_\ast(M))$ if we send the effective value of $\alpha$ to $+1$; hence, from (\ref{points}), we have
$\chi(HF^{w=0}_\ast(M)) =  \sum_{x} \, \pm 1$,  where the $x$'s are the isolated points that span the zero-dimensional space ${\cal M}_f$ of flat (ordinary) $SU(2)$-connections on $M$; this just reflects the established fact that $\chi(HF^{w=0}_\ast(M))$ can also be  interpreted as the Euler number $\chi({\cal M}_f)$~\cite{blau}. By these two points, (\ref{chi=chi}) will then mean that $\chi({\cal M}_f)$ is zero if $M$ has positive scalar curvature -- in other words, if $M$ has \emph{positive} scalar curvature,  ${\cal M}_f$ is either empty or spanned by an \emph{equal} number of positively and negatively-oriented points.  

 \bigskip\noindent{\it Implications For The Seiberg-Witten Invariants Of $M$}
 
Another useful thing to note is that it was pointed out in~\cite{moore-marino} that $\lambda_{CWL} (M)$ in (\ref{cwl}) can be expressed as a sum of all coefficients of the Reidemeister-Milnor torsion; in turn, by the work of Meng-Taubes in~\cite{MT}, this sum is given by a certain combination of SW invariants of $M$ called the SW series $SW(t_i)$, where the $t_i$'s are variables whose details will not be important. Consequently, by (\ref{cwl}) and (\ref{prt}), we have   
\be
SW(M, \pi({\hat {\frak s}})) = \sum_{x \in H} \sum_{{\frak s}_M \vert {\bar c}_1({\frak s}_M) = x} SW (M, {\frak s}_M)
\label{SWseries}
\ee
up to a sign, where $H = H^2(M, \mathbb Z) / {\rm Tor}(H^2(M, \mathbb Z))$ is the torsion-free part of the second integral cohomology of $M$, ${\bar c}_1({\frak s}_M)$ is the projection of $c_1({\frak s}_M)$ to $H$, and $c_1(\pi({\hat {\frak s}})) \in H$. 
 
Once again, we can validate (\ref{SWseries}) for $M = \Sigma_g \times {\bf S}^1$ (or $X = \Sigma_g \times {\bf T}^2$) with $g \geq 1$. As we saw in $\S$5.4, the magnitude of ${\rm Gr}(c_1({\cal E}))$ and thus that of $SW(X, {\hat {\frak s}}) = SW(M, \pi({\hat {\frak s}}))$ on the LHS of (\ref{SWseries}) is equal to $1$ and $0$ for $g=1$ and $g > 1$, respectively. At the same time, it is known that the SW series and hence the RHS of (\ref{SWseries}) is given by $SW(1)$, where $SW(t) = (t^{1/2} - t^{-1/2})^{2g -2}$; in other words, the magnitude of the RHS of (\ref{SWseries}) is $1$ and $0$ for $g=1$ and $g> 1$, too. Therefore, for  $M = \Sigma_g \times {\bf S}^1$ with $g \geq 1$,  (\ref{SWseries}) is found to be consistent with all known mathematical results.

 \bigskip\noindent{\it A Non-Vanishing Theorem For  The Monopole Floer Homology Of $M$}

Now consider $X = M \times {\bf S}^1$ to be \emph{general} with $b^+_2(X) = b_1(M) > 1$, where $M$ is a compact, oriented three-manifold. Let $\Sigma$ be an oriented two-surface of  genus $g > 0$ that is smoothly-embedded in $M$; then, the normal bundle of $\Sigma$ in $X$ is trivial,\footnote{Consider the restriction $TM\vert_{\Sigma}$  to $\Sigma$ of the tangent bundle $TM$ of $M$; it splits as $TM\vert_\Sigma = T\Sigma \oplus TN$, where $T\Sigma$ and $TN$ are the tangent and normal bundles of $\Sigma$ in $M$, respectively. Note that $TM$ is oriented and so is $T\Sigma$; hence, $TN$ is also orientable. However an orientable real line bundle such as $TN$ must be trivial; therefore, the normal bundle of $\Sigma$ in $X$ is also trivial, and it is given by $\Sigma \times {\bf R}^2$.} i.e., $\Sigma \cap \Sigma =0$. As such, by $SW(X, \pi^{-1}({\frak s}_M)) = SW(M, {\frak s}_M)$, and by Theorem 1.3 of~\cite{OS 2} -- which generalizes (\ref{cor 1.7}) to $X$ with $b_1 \neq 0$ -- we have 
\be
\vert \langle c_1( {\frak s}_M), [\Sigma] \rangle \vert  \leq (2g-2)
\label{OS 2nd paper}
\ee
if $SW(M, {\frak s}_M) \neq 0$. Hence, since $\chi (HM_\ast(M, {\frak s}_M)) = SW(M, {\frak s}_M)$, it will mean that 
\be
HM_\ast(M, {\frak s}_M) \neq 0
\ee
as long as (\ref{OS 2nd paper}) holds. For $c_1({\frak s}_M)$ non-torsion, this claim is just Corollary 40.1.2 of~\cite{3-manifolds}. 


\def\mAB{{\mathscr A}'_B}

\newsubsection{``Ramified'' Generalizations Of Various Relations Between Donaldson And Floer Theory}

We shall now formulate, purely physically,  ``ramified'' generalizations of various formulas presented by Donaldson and Atiyah in~\cite{Donald, Weyl} that relate ordinary Donaldson and Floer theory on four-manifolds with boundaries.

\bigskip\noindent{\it ``Ramified'' Donaldson Invariants With Values In Knot Homology Groups From ``Ramified'' Instantons}

To this end, let general $X = B \times {\bf R}_{\geq 0}$, where $B$ can be interpreted as the boundary of $X$, and the half real-line ${\bf R}_{\geq 0}$ can be interpreted as the ``time'' direction. Let the surface operator $\Sigma = K_B \times {\bf R}_{\geq 0}$, where $K_B$ is an arbitrary knot embedded in $B$. In such a case, there exists in the supersymmetry algebra a Hamiltonian $H$ which generates translations along ${\bf R}_{\geq 0}$, and by replacing ${\bf S}^1$ with ${\bf R}_{\geq 0}$ in our earlier explanation, we find that the ``ramified'' Donaldson-Witten theory  can be interpreted as a supersymmetric quantum mechanical sigma-model with worldline ${\bf R}_{\geq 0}$, target manifold $\mAB / {\cal G}'_B$ -- the space of all gauge-inequivalent classes of ``ramified'' $SU(2)$-connections $\mAB$ on $B$,  and potential $h_B = \half \int_{B} {\rm Tr} \, (\mAB \wedge d\mAB + {2\over 3} \mAB \wedge \mAB \wedge \mAB)$ -- the Chern-Simons functional of $\mAB$. In particular, $\mAB$ can be regarded a gauge connection of an $SU(2)$-bundle over $B \backslash K_B$ whose holonomy around the meridian of $K_B$ is given by ${\rm exp} (2 \pi i \alpha)$, while the $\cq$-cohomology of the sigma-model -- which is furnished by the supersymmetric ground states that correspond to the critical points of $h_B$ -- can, in fact, be identified with the ``ramified'' instanton Floer homology $HF_\ast(B; K_B; \alpha)$.

According to the general ideas of quantum field theory, when the theory is formulated on such an $X$,  one must specify the boundary values of the path-integral fields along $B$. Let us denote $\Phi_B$ to be the restriction of these fields to $B$; then, in the space $\mathscr H$ of functionals of the $\Phi_B$, specifying a set of boundary values for the fields on $B$ is tantamount to selecting a functional $\Psi(\Phi_B) \in \mathscr H$. Since the $\cq$-cohomology of the sigma-model is annihilated by $H$, i.e., it is time-invariant,  one can take an arbitrary time-slice in $X$ and study the quantum theory formulated on $B$ instead; in this way, $\Psi(\Phi_B) \in \mathscr H$ can be interpreted as a state in the Hilbert space $\mathscr H$ of the quantum theory on $B$. As a result, via a state-operator mapping of the topological field theory, the correlation function ``with boundary values of the fields determined by $\Psi$'' will be given by 
\be
\langle \mo_1 \dots \mo_n \rangle_{\Psi(\Phi_B)} = \int \CD \Phi \  e^{- {S_E}} \ \mo_1 \dots \mo_n \cdot \Psi(\Phi_B).
\label{CF_b}
\ee

Since the theory ought to remain topological in the presence of a boundary $B$, according to our discussion in $\S$2.3, it must be that $\{\cq, \mo_i] = 0 = \{\cq, \Psi]$. Moreover, if $\Psi = \{\cq, \dots]$, the fact that $\{\cq, \mo_i] = 0$ implies that (\ref{CF_b}) will also be zero. Thus, (\ref{CF_b})  depends on $\Psi$ via its interpretation as a $\cq$-cohomology class only, and since $\Psi$ is associated with the quantum theory on $B$, we can identify $\Psi$ as a class in the ``ramified'' instanton Floer homology $HF_\ast(B; K_B; \alpha)$. Altogether, since the $\mo_i$'s represent either the operators $I'_0(p)$ or $I'_2(S)$ in (\ref{I's zero-modes 1})-(\ref{I's zero-modes}), by (\ref{correlation function 1}), we find that (\ref{CF_b}) will represent a ``ramified'' Donaldson invariant with values in $HF_\ast(B; K_B; \alpha)$ -- a knot homology group from ``ramified'' instantons. This is just a ``ramified'' generalization of the ordinary relation between Donaldson and Floer theory on $X$ described by Donaldson in~\cite{Donald}. 

\bigskip\noindent{\it Interpretation As A Scattering Amplitude Of ``Three-One Branes''}

Now, let us assume that the total boundary $\partial X$ of $X$ consists not of a single boundary $B$, but a disjoint union of boundaries $B_j$, $j =1, \dots, r$; i.e., 
\be
\partial X = \bigsqcup^r_{j =1} B_j.
\label{boundary}
\ee
Let the surface operator $\Sigma = K_{\partial X} \times (X \backslash \partial X)$. If one is to choose the $\Psi(\Phi_{B_j})$'s appropriately such that one can replace all the $\mo_i$'s with the identity operator $1$ and yet have a non-vanishing path-integral,  the resulting correlation function
\be
\langle 1 \rangle_{\Psi(\Phi_{B_1});\dots;\Psi(\Phi_{B_r})} = \int \CD \Phi \  e^{- {S_E}} \ \Psi(\Phi_{B_1}) \dots \Psi(\Phi_{B_r}) 
\label{scattering}
\ee
can be interpreted as a scattering amplitude of incoming and outgoing ``three-one branes'' (the knot $K_{B_j}$ being the one-brane with ``magnetic'' charge $\alpha$ that is embedded in the three-brane $B_j$). 

At any rate, according to the general ideas of quantum field theory, one can also write
\be 
\langle 1 \rangle_{\Psi(\Phi_{B_1});\dots;\Psi(\Phi_{B_r})} = \int \CD \Phi \  e^{- {S_E}} \ \Psi(\Phi_{\partial X}).
\ee
As such,  (\ref{scattering}) can be expressed as
\be
\int \CD \Phi \  e^{- {S_E}} \ \Psi(\Phi_{\partial X})  = \int \CD \Phi \  e^{- {S_E}} \ \Psi(\Phi_{B_1}) \dots \Psi(\Phi_{B_r}).
\ee
In turn, this implies the relation
\be
HF_\ast(\partial X; K_{\partial X}; \alpha) = HF_\ast(B_1; K_{B_1}; \alpha) \otimes HF_\ast(B_2; K_{B_2}; \alpha) \otimes \dots \otimes HF_\ast(B_r; K_{B_r}; \alpha)
\label{tensor}
\ee
 for knot homology groups from ``ramified'' instantons -- which can be interpreted as a ``ramified'' generalization of eqn. (6.5) of~\cite{Donald} -- that describes a scattering amplitude of  ``three-one branes''.

\bigskip\noindent{\it A Poincar\'e Duality Map Of Knot Homology Groups From ``Ramified'' Instantons}

Note that it is known~\cite{freedman} that one can always decompose a general $X$ along a homology three-sphere $Y$ into two parts $X^+$ and $X^-$, as shown in fig.~1 below. Let $\Sigma^{\pm}$ be the parts of the surface operator $\Sigma$ which are embedded in $X^\pm$, and let $\Sigma^+_K = K_Y$ and $\Sigma^-_{K} = K_{\overline Y}$ be their corresponding knot components embedded in $Y$ and $\overline Y$, respectively, where $\overline Y$ and $K_{\overline Y}$ are oppositely-oriented copies of $Y$ and $K_Y$.

Now, consider the sector of the theory with ``ramified'' instanton number $k'$ given in (\ref{k'}), i.e., the $N_\psi = 0$ sector. The relevant non-vanishing observable is then the partition function $\pf$. Since $X = X^+ \cup_Y X^-$, according to the general ideas of quantum field theory, one can evaluate $\pf$ as a path-integral over $X^+$\emph{ towards} $Y$ followed by a second path-integral over $X^-$ \emph{away} from $\overline Y$. Specifically, an independent path-integral over $X^+$ towards $Y$ will determine a state $\langle + |$ in the Hilbert space $\mathscr H_Y$ of the quantum theory on $Y$, while an independent path-integral over $X^-$ away from $\overline Y$ will determine a state $| - \rangle$ in the Hilbert space $\mathscr H_{\overline Y}$ of the quantum theory on $\overline Y$. As $\mathscr H_{\overline Y}$ is canonically the dual of $\mathscr H_{Y}$, we have 
\be
\langle + | - \rangle = \pf.
\label{map}
\ee

According to our above discussions, the independent path-integral over $X^+$ will be given by 
\be
\langle + | = \int \CD \Phi \  e^{- {S_E}} \ \Psi(\Phi_{Y}), 
\label{+}
\ee
while the independent path-integral over $X^-$ will be given by
\be
| - \rangle = \int \CD \Phi \  e^{- {S_E}} \ \Psi(\Phi_{\overline Y}),
\label{-}
\ee
where $\Psi(\Phi_{Y})$ and $\Psi(\Phi_{\overline Y})$ can be interpreted as classes in $HF_\ast(Y; K_Y; \alpha)$ and $HF_\ast(\overline Y; K_{\overline Y}; \alpha)$, respectively. At the same time,  as explained in $\S$4, we have $\pf \in \mathbb Z$. Altogether, via (\ref{+}) and (\ref{-}),  one can interpret (\ref{map}) as a Poincar\'e duality map 
\be
HF_\ast(Y; K_Y; \alpha) \otimes HF_\ast(\overline Y; K_{\overline Y}; \alpha) \rightarrow \mathbb Z
\label{PD map}
\ee
of knot homology groups from ``ramified'' instantons. Note that (\ref{PD map}) can be interpreted as a ``ramified'' generalization of a Poincare duality map of ordinary instanton Floer homology groups described by Atiyah in~\cite{Weyl}.     

\begin{figure}
  \centering
    \includegraphics[width=0.8\textwidth]{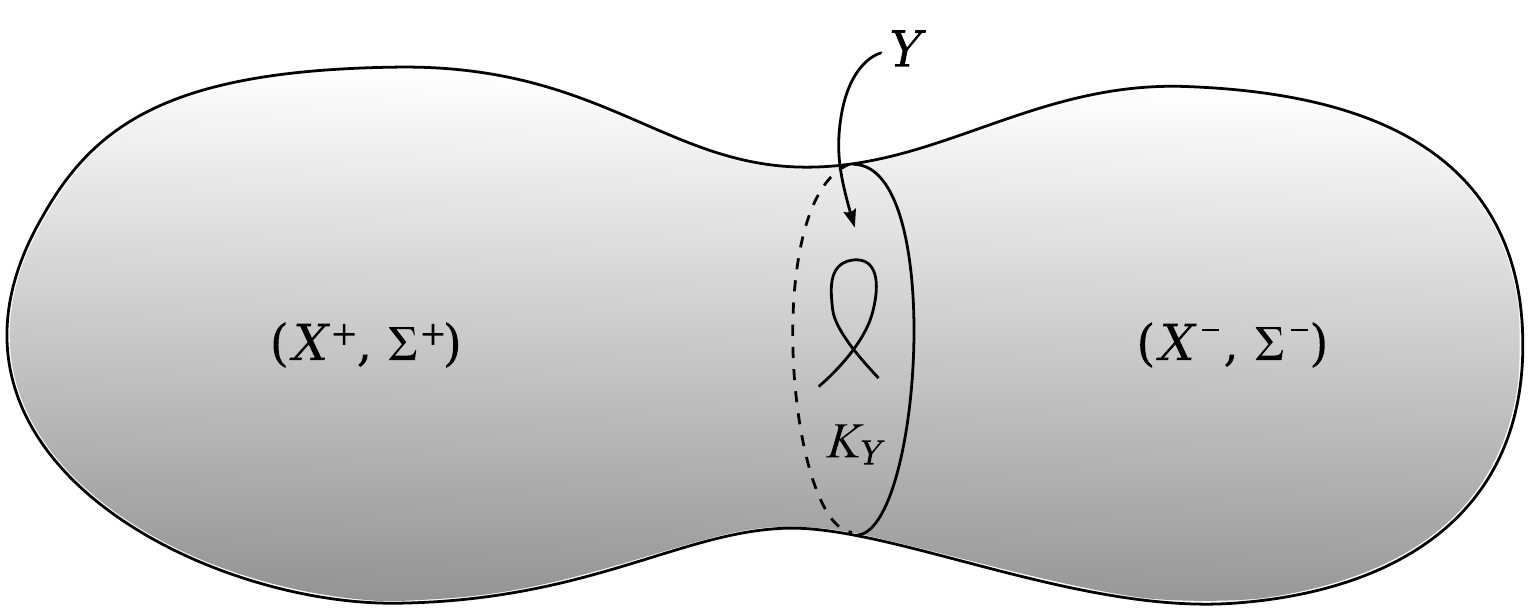}
  \caption{$X = X^+ \cup_Y X^-$}
\end{figure}

\newsection{Generalization Involving Multiple Surface Operators}

Notice that our physical derivation of Taubes' result in $\S$4 involved only a single, connected surface operator $\Sigma$ that is nontrivially-embedded in $X$ with $\Sigma \cap \Sigma > 0$. Let us now revisit that section and consider the case where one has multiple surface operators which are nevertheless similar to $\Sigma$.

\bigskip\noindent{\it A Disconnected Pseudo-Holomorphic Curve}

Let us start by considering the ``total'' surface operator 
\be
e = \sum_m \, \Sigma_m,
\ee
where $\Sigma_m$ -- like $\Sigma$ characterized earlier by (\ref{adj Taubes}) and (\ref{PD sigma}) -- is selected from the $b^+_2$ number of homology cycles in the basis $\{U_i\}_{i =1, \dots, b_2}$ which has a purely diagonal, unimodular intersection matrix. In particular, we have
\be
\Sigma_i \cap \Sigma_j = 0 \quad {\rm when} \quad i \neq j,
\label{intersect}
\ee
and therefore, the ``total'' surface operator $e$ consists of disjoint, non-multiply-covered components furnished by the ``member'' surface operators $\Sigma_m$ which are themselves connected pseudo-holomorphic curves in $X$. Consequently, $e$ is also a pseudo-holomorphic curve, albeit a disconnected one, and since the Poincar\'e duals of the $\Sigma_m$'s are such that
\be
\delta_{\Sigma_m} = \delta^+_{\Sigma_m}
\label{delta}
\ee
for all $m$, it will mean that
\be
e \cap e > 0. 
\ee

\bigskip\noindent{\it The Corresponding Moduli Space Of ``Ramified'' $SU(2)$-Instantons}

With the insertion of multiple disjoint surface operators as represented by the ``total'' surface operator $e$, the path-integral of the topological $SU(2)$ gauge theory localizes onto supersymmetric configurations which satisfy (cf.~(\ref{delta})) 
\be
F^+_e =  2 \pi i  \sum_m \alpha_m  \delta_{\Sigma_m},
\label{ram total}
\ee
where $\alpha_m$ is the ``classical'' parameter of the corresponding ``member'' surface operator $\Sigma_m$, and $F^+_e$ can be interpreted as an imaginary-valued curvature two-form of some complex line bundle with a self-dual ${\frak u}(1)$-valued connection $A_e$.  

Since the surface operators are disjoint,  the holonomies of the gauge field around small circles linking the various $\Sigma_m$'s will not ``mix'' with one another. As such, one can rewrite (\ref{ram total}) as a set of relations
\be
F^+_m = 2 \pi i  \alpha_m  \delta_{\Sigma_m}, \qquad {\rm for} \quad  m = 1, 2, \dots, 
\label{ram member}
\ee
where $F^+_m$ can be interpreted as an imaginary-valued curvature two-form of some complex line bundle with a self-dual ${\frak u}(1)$-valued connection $A_m$. 

In other words, the path-integral localizes onto the moduli space $\CM'$ of ``ramified'' $SU(2)$-instantons spanned by field configurations which satisfy the relation (\ref{ram total}); the equivalent relations in (\ref{ram member}) then imply that $\CM'$ can actually be expressed as
\be
\CM' = \bigotimes_m \CM'_m,
\label{M'}
\ee 
where $\CM'_m$ is the moduli space spanned by field configurations which satisfy the $m$-th relation in (\ref{ram member}). 

Similar to the case of a single surface operator, $N_{\psi}$ is given by the expression  
\be
N_{\psi} = 8 k' - {3 \over 2} (\chi + \sigma), 
\ee
although now, the ``ramified'' instanton number $k'$ is given by
\be
k' = k + 2 \sum_m \alpha_m l_m - \sum_m \alpha_m^2  (\Sigma_m \cap \Sigma_m).
\label{k' = k + l}
\ee
In particular, $k = \int_X c_2(E)$ and $l_m = - \int_{\Sigma_m} c_1({L})$, and according to our discussion surrounding (\ref{c=c}), the value of $k$ -- for any particular choice of $X$ and set of surface operators with parameters $\{ \alpha_m\}$ and positive self-intersection numbers $\{ \Sigma_m \cap \Sigma_m \}$ -- determines the values of all the $l_m$'s,  and vice-versa. Thus, from (\ref{k' = k + l}),  we find that the value of $k'$ is in one-to-one correspondence with the set $\{l_m\}$.  

\bigskip\noindent{\it The $N_{\psi} = 0$ Sector}

Let us now consider the sector of the $SU(2)$ theory where $k'$ is as given in (\ref{k'}); i.e., the sector where $N_{\psi} = {\rm dim} (\CM') = 0$.  
Since electric-magnetic duality in the low-energy $U(1)$ theory implies that there is a one-to-one correspondence between $c_1(L)$ and $\lambda$, there is, according to our preceding discussion, a one-to-one correspondence between $k'$ and ${\frak s} = -i \lambda$ for any particular choice of $X$ and set of surface operators.  Then, by sending the effective value of $\alpha_m$ to $+1$ for every $m$ whence all the surface operators become ``ordinary'',  and by repeating the arguments behind (\ref{0-instanton})-(\ref{H}) whilst noting the fact that  if $\CM'$ of (\ref{M'}) is zero-dimensional, so are the spaces $\CM'_m$, we get
\be
SW ({\frak s}) =  \sum_x q(x),
\label{gen SW}
\ee 
where
\be
q(x) = \prod_m \, {\rm sign} \, ({\rm det} \, {\cal D}_m).
\label{gen 1}
\ee
Here, the $x$'s are the points which span the zero-dimensional space $H_e$ of solutions to the relation
\be
c_1({\cal E}_e) = \delta_e,
\label{gen 2}
\ee 
where ${\cal E}_e$ is a nontrivial complex line bundle with a self-dual ${\frak u}(1)$-valued connection $A_e$ whence $c_1({\cal E}_e) \cdot c_1({\cal E}_e) > 0$; ${\cal D}_m$ for all $m$ is a first-order elliptic operator whose kernel is the tangent space to the space $H_m$ of solutions to the relation 
\be
c_1({\cal E}_m) = \delta_{\Sigma_m},
\label{gen 3}
\ee
where ${\cal E}_m$ is a nontrivial complex line bundle with a self-dual ${\frak u}(1)$-valued connection $A_m$ whence $c_1({\cal E}_m) \cdot c_1({\cal E}_m) > 0$; and 
\be
{\cal E}_e = \otimes_m {\cal E}_m.
\label{gen 4}
\ee 

Note that (\ref{gen 1})-(\ref{gen 4}) mean that one can rewrite (\ref{gen SW}) as
\be
SW ({\frak s}) =  {\rm Gr}(c_1({\cal E}_e)),
\label{gen 5}
\ee
where ${\rm Gr}(c_1({\cal E}_e))$ is the Gromov-Taubes invariant defined in~\cite{CT 1} for a disconnected, non-multiply-covered, pseudo-holomorphic curve $e$ whose fundamental class is Poincar\'e dual to $c_1({\cal E}_e)$.  

\bigskip\noindent{\it Arriving At Taubes' Result}

Before we proceed further, note that the analysis carried out in $\S$3.2 can be generalized to the present case with multiple disjoint surface operators: one simply replaces ``$\alpha \delta_{\Sigma}$'' in the relevant analysis therein with ``$\sum_m \alpha_m \delta_{\Sigma_m}$''. With this in mind, note that   since (\ref{gen 5}) is valid for $X$ of SW simple-type, i.e., $\lambda^2 - (2 \chi + 3 \sigma) /4 = d_{e} = -c_1(L^2_d) [e] + e \cap e = 0$, we find that the generalizations of (\ref{SW = SW}), (\ref{exp SW}) and ${\frak s'} = \frak s - \delta_e$ will imply that the LHS of (\ref{gen 5}) is $SW(\frak s - \delta_e)$. In fact, since the non-vanishing contributions to the RHS of (\ref{gen 5}) localize around supersymmetric field configurations which obey (\ref{gen 2}), the LHS of  (\ref{gen 5}) can actually be written as $SW(\frak s - c_1({\cal E}_e))$;  as a result, we have
\be
SW(\frak s - c_1({\cal E}_e)) =   {\rm Gr}(c_1({\cal E}_e)). 
\label{taubes gen}
\ee 

As each $\Sigma_m$ is a connected pseudo-holomorphic curve with positive self-intersection, it must satisfy (\ref{inequality new}) and (\ref{adj Taubes}) simultaneously. This implies that we necessarily have ${\frak s} = {1 \over 2}c_1(L^2_d) = {1 \over 2}c_1(K)$, as in the case of a single surface operator. By noting that as $b^+_2 > 1$, the ordinary SW invariants satisfy $SW( \bar {\frak s}) = \pm SW( - \bar {\frak s})$ for any ordinary $\spin$-structure $\bar {\frak s}$~\cite{monopoles}, we can also write (\ref{taubes gen}) as
\be
SW (\hat {\frak s}_e) = \pm  {\rm Gr}(c_1({\cal E}_e)),
\label{prt gen}
\ee
where 
\be
\hat {\frak s}_e = {1 \over 2} c_1({\mathscr L}_e),
\label{prt 1 gen}
\ee
and 
\be
{\mathscr L}_e = K^{-1} \otimes {\cal E}^2_e. 
\label{prt 2 gen}
\ee
Moreover, since $c_1(L^2_d) = c_1(K)$, by (\ref{intersect}) and (\ref{gen 4}), we find that the condition $d_{e} =0$  can also be expressed as
\be
d_e = \sum_m d_m = 0,
\label{dm = d}
\ee
where
\be
d_e = - c_1(K) \cdot c_1({\cal E}_e) +   c_1({\cal E}_e) \cdot  c_1({\cal E}_e), 
\label{d condition gen}
\ee
and
\be
d_m =  - c_1(K) \cdot c_1({\cal E}_m) +   c_1({\cal E}_m) \cdot  c_1({\cal E}_m).
\label{dm}
\ee
Because (\ref{d condition gen}) and (\ref{dm}) are the \emph{non-negative} dimensions of $H_e$ and $H_m$ as defined mathematically in~\cite{CT ps}, we see that (\ref{dm = d}) is indeed consistent with the fact that  $H_e$ and therefore all the $H_m$'s are zero-dimensional as implied by $N_{\psi} = 0$. In turn, the fact that $d_m =0$ implies, via (\ref{dm}) and (\ref{adj Taubes}), that the genus $g_m$ of the ``member'' pseudo-holomorphic curve represented by $c_1({\cal E}_m)$ will be given by 
\be
g_m = 1 + c_1({\cal E}_m) \cdot c_1({\cal E}_m). 
\label{gps gen}
\ee

Finally, note that (\ref{prt gen})-(\ref{gps gen}) are \emph{precisely} Taubes' theorem~\cite{CT 1} equating the ordinary Seiberg-Witten invariants to the Gromov-Taubes invariants for disconnected curves in $X$!  This implies that the novel mathematical identities obtained in the previous section can be generalized to hold for \emph{disconnected}, non-multiply-covered,  pseudo-holomorphic curves in $X$ with positive self-intersection, too. Nevertheless, in favor of brevity, we will not verify this explicitly. 

\def\cC{{\cal C}}
\def\cE{{\cal E}}

\newsection{Further Application Of Our Physical Insights And Results}

Let us now, in this final section,  apply some of our physical insights and results obtained hitherto to 1).~elucidate certain key properties of the knot homology groups from ``ramified'' instantons discussed in $\S$5.3 and $\S$5.6; 2).~tell us more about the monopole Floer homology groups discussed in $\S$5.5; 3).~tell us more about the ordinary Seiberg-Witten invariants of a compact, oriented, symplectic four-manifold  with $b_1 =0$ and $b^+_2 > 1$.

\newsubsection{Properties Of Knot Homology Groups From ``Ramified'' Instantons}

\medskip\noindent{\it Metric-Independence}

Let us consider a decomposition of a general $X$ along two disjoint, compact, connected, oriented three-manifolds $Y_0$ and $Y_1$ into three parts $X^+$, $X^-$ and $X'$, as shown in fig.~2 below.  Let $\Sigma^\pm$ and $\Sigma'$ be the parts of the surface operator $\Sigma$ which are embedded  in $X^\pm$ and $X'$,  and let $\Sigma^+_{K} = K_{Y_0}$, $\Sigma^-_{K}  = K_{{\overline Y}_0} \cup K_{Y_1}$ and $\Sigma'_{K} = K_{Y_1}$ be their corresponding knot components embedded in $Y_0$, ${\overline Y}_0$ and $Y_1$,  respectively, where $K_B$ indicates the oriented knot embedded in the oriented three-manifold $B$ such that the holonomy of the gauge field around its meridian is ${\rm exp}(2 \pi i \alpha)$, and  $K_{\overline B}$ and $\overline B$ are just oppositely-oriented copies of $K_B$ and $B$.

Let us now, as was done for a similar case in $\S$5.6, compute the path-integral over the middle segment labeled by $X^-$. In this case, there will be two sets of boundary values of the fields: one at ${\overline Y}_0$, and the other at $Y_1$. As such, according to the general ideas in quantum field theory, the path-integral will be given by
\be
\Psi(\Phi') = \int_{\Phi_{Y_1} = \Phi'} \CD \Phi \  e^{- {S_E}} \ \Psi(\Phi_{{\overline Y}_0}).
\label{pi}
\ee
An explanation of the above formula is in order.  Firstly, $\Phi_B$ indicates the restriction of the path-integral fields to $B$; correspondingly, $\Psi(\Phi_B)$ is a functional of $\Phi_B$ which determines the boundary values of the fields on $B$. Secondly, the path-integral is computed over all fields $\Phi$ which when restricted to $Y_1$, have values $\Phi'$; this takes care of the boundary values on $Y_1$;  according to our discussions in $\S$5.6, the insertion of the operator $\Psi(\Phi_{{\overline Y}_0})$ will then take care of the boundary values on ${\overline Y}_0$. Thirdly,  we have assumed the boundary values of the fields on ${\overline Y}_0$ and therefore  $\Psi(\Phi_{{\overline Y}_0})$,  to be  a priori determined, while on the other hand,  we have assumed the boundary values of the fields on $Y_1$ and therefore $\Phi'$, to be a priori \emph{undetermined}. As such, the path-integral will depend on $\Phi'$, and therefore, it can also be interpreted as a functional $\Psi(\Phi')$ of $\Phi'$, as written in (\ref{pi}).  

\begin{figure}
  \centering
    \includegraphics[width=0.8\textwidth]{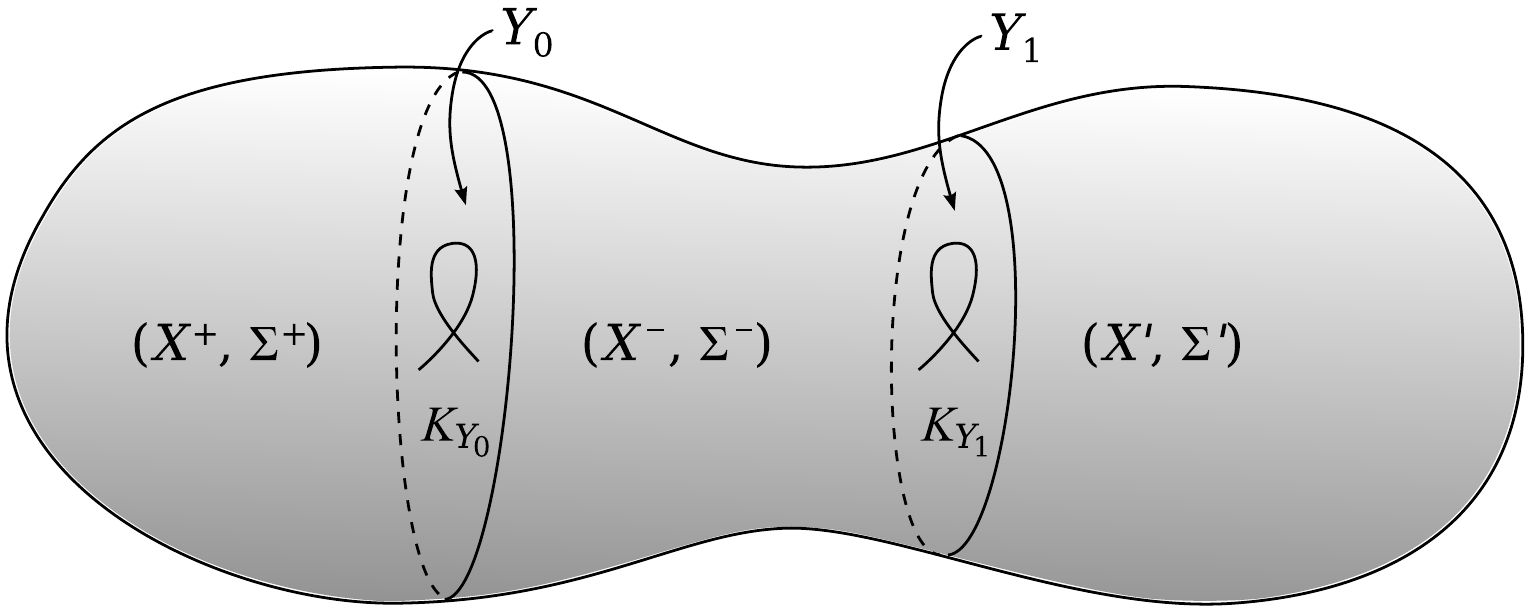}
  \caption{$X = X^+ \cup_{Y_0} X^- \cup_{Y_1} X'$}
\end{figure}

Note at this point that the integration measure $\CD \Phi$ is invariant under supersymmetry; in other words, it is $\cq$-closed. Recall also from (\ref{S}) that $S_E = \{\cq, \dots \}$ (since, as explained in footnote~13, we are considering surface operators with $\eta =0$, while the $\Theta$-angle can always be set to zero via an irrelevant chiral rotation of the massless fermions). Altogether, since $\cq^2 =0$, it will mean that  if $\{\cq, \Psi(\Phi_{{\overline Y}_0})] = 0$, then $\{\cq, \Psi(\Phi') ] =0$, and if $\Psi(\Phi_{{\overline Y}_0}) = \{ \cq, \dots ]$, then $\Psi(\Phi') = \{ \cq, \dots ]$, too. Therefore, (\ref{pi}) represents a map ${\cal H} : \Psi(\Phi_{{\overline Y}_0}) \to \Psi(\Phi')$ of $\cq$-cohomology classes. In addition, as explained in $\S$2.3, due to the stress-tensor $T_{\mu \nu}$ of the underlying physical theory being $\cq$-exact,  $\cal H$ is necessarily invariant under metric deformations of $X$.

Now, let us decompose $X$ along three disjoint, compact, connected, oriented three-manifolds $Y_0$, $Y_1$ and $Y_2$ into four parts $X^+$, $X^-$, $X'$ and $X''$, as shown in fig.~3 below.  Let $\Sigma^\pm$, $\Sigma'$ and $\Sigma''$ be the components of the surface operator $\Sigma$ which are embedded  in $X^\pm$, $X'$ and $X''$,  and let $\Sigma^+_{K} = K_{Y_0}$, $\Sigma^-_{K}  = K_{{\overline Y}_0} \cup K_{Y_1}$, $\Sigma'_K = K_{{\overline Y}_1} \cup K_{Y_2}$ and $\Sigma''_{K} = K_{{\overline Y}_2}$ be their corresponding knot components embedded in $Y_0$, $Y_1$, $Y_2$ and their oppositely-oriented copies, respectively.    What we would like to do next is to compute the path-integral over the region $X^- \cup_{Y_1} X'$. If we assume the boundary values of the fields on $Y_2$ -- like those on  ${\overline Y}_0$ -- to be a priori determined, the path-integral will be given by
\be
Z(X^- \cup_{Y_1} X')  = \int_{\Phi_{Y_1} = \Phi'} \CD \Phi \  e^{- {S_E}} \ \Psi(\Phi_{{Y}_2}) \cdot  \Psi(\Phi_{{\overline Y}_0}).
\label{Zxx}
\ee
That being said, according to the general ideas of quantum field theory,  one can also compute $Z(X^- \cup_{Y_1} X')$ as a path-integral over $X^-$ -- away from ${\overline Y}_0$ and towards $Y_1$ -- followed by a second path-integral over $X'$ -- away from ${\overline Y}_1$ and towards $Y_2$.  Thus, from our above discussion leading to (\ref{pi}), and (\ref{Zxx}), we can write
\be
  \int_{\Phi_{{Y}_1} = {\Phi}'} \CD \Phi \  e^{- {S_E}} \ \Psi(\Phi_{{Y}_2}) \cdot  \int_{\Phi_{Y_1} = \Phi'} \CD \Phi \  e^{- {S_E}} \ \Psi(\Phi_{{\overline Y}_0}) = \int_{\Phi_{{Y}_1} = {\Phi}'} \CD \Phi \  e^{- {S_E}} \  \Psi(\Phi_{{Y}_2}) \cdot \Psi(\Phi_{{\overline Y}_0}),
\label{homo}
\ee 
where we have made use of the fact that specifying the a priori undetermined boundary values of the fields on ${\overline Y}_1$ is equivalent to specifying those on its mirror ${Y}_1$. Notice that (\ref{homo}) means that 
\be
 {\cal H} ( \Psi(\Phi_{{Y}_2})) \cdot {\cal H} ( \Psi(\Phi_{{\overline Y}_0}))=  {\cal H} ( \Psi(\Phi_{{Y}_2}) \cdot \Psi(\Phi_{{\overline Y}_0})),
\ee
i.e., the map $\cal H$ is a \emph{homomorphism}. 

\begin{figure}
  \centering
    \includegraphics[width=0.8\textwidth]{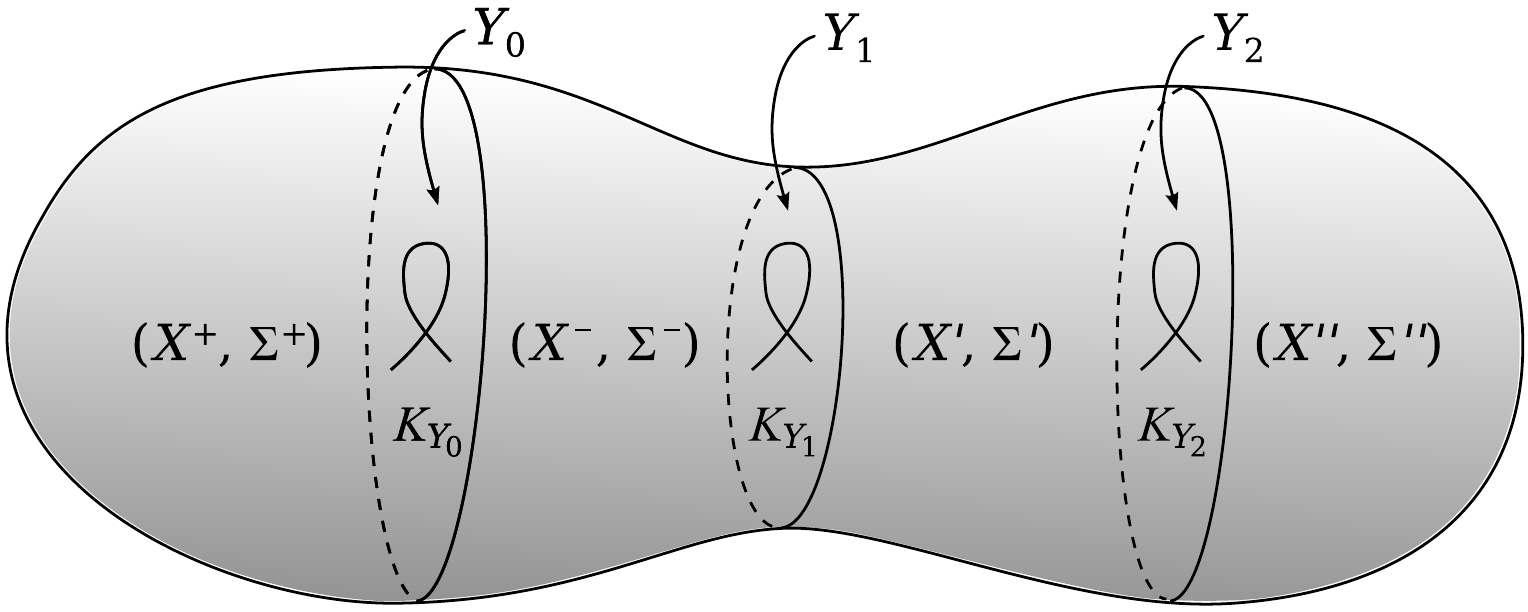}
  \caption{$X = X^+ \cup_{Y_0} X^- \cup_{Y_1} X'  \cup_{Y_2} X''$}
\end{figure}

As per our discussions in $\S$5.6, we find that $\Psi(\Phi_{{\overline Y}_0})$, $\Psi(\Phi_{Y_1})$ and $\Psi(\Phi_{Y_2})$ will correspond  to classes in $HF_\ast({\overline Y}_0; K_{{\overline Y}_0}; \alpha)$,  $HF_\ast(Y_1; K_{Y_1}; \alpha)$ and $HF_\ast(Y_2; K_{Y_2}; \alpha)$, respectively. Also, according to our discussions in $\S$5.6, the state $\Psi(\Phi_{{\overline Y}_0}) \in {\mathscr H}_{{\overline Y}_0}$ is in fact \emph{dual} to the state $\Psi(\Phi_{{Y}_0}) \in {\mathscr H}_{{Y}_0}$ (where ${\mathscr H}_B$ refers to the Hilbert space of the quantum theory on $B$), i.e., we can identify $HF_\ast({\overline Y}_0; K_{{\overline Y}_0}; \alpha)$ with 
$HF_\ast({Y}_0; K_{{Y}_0}; \alpha)$. Hence,  the map (\ref{pi}) can also be interpreted as the following homomorphism
\be
{\cal H}:  HF_\ast({Y}_0; K_{{Y}_0}; \alpha) \rightarrow HF_\ast(Y_1; K_{Y_1}; \alpha)
\label{homo H}
\ee 
on knot homology groups from ``ramified'' instantons.    Moreover, $X^-$ is a connected, oriented manifold-with-boundary, and it contains a properly embedded oriented surface-with-boundary $\Sigma^-$,  whence  we have an orientation-preserving diffeomorphism of pairs
\be
r : ({\overline Y}_0, K_{{\overline Y}_0}) \cup (Y_1, K_{Y_1}) \rightarrow (\partial X^-, \partial \Sigma^-).
\ee
In other words, we have a cobordism  from $({Y}_0, K_{{Y}_0})$ to $(Y_1, K_{Y_1})$ -- that underlies the definition of the path-integral over $X^-$ -- which  gives rise to the homomorphism $\cal H$ of (\ref{homo H});  since $\cal H$  is invariant under metric deformations of $X$, it will depend only on the diffeomorphism class of the cobordism, albeit up to a sign; this sign is determined by a choice of the zero-modes of $(A'_\mu, \psi_\mu, \chi^+_{\mu\nu})$  in the integration measure of (\ref{pi}), i.e., a choice of orientation for the line $\Lambda^{\rm max}H^1(X^-; \mathbb R) \otimes \Lambda^{\rm max}H^{2,+}(X^-; \mathbb R) \otimes \Lambda^{\rm max}H^1(Y_1; \mathbb R)$.    This result has also been proved via a distinct mathematical approach by Kronheimer and Mrowka as Proposition 3.27 in~\cite{dym}.  In turn, this means that $HF_\ast(M; K; \alpha)$ for some compact, connected, oriented three-manifold $M$ with an oriented knot $K$ embedded in it, will be independent of the metric on $M$. 

\bigskip\noindent{\it The Identity Map} 

Let us consider the setup in fig.~2 again.  If $Y_1 = Y_0$ and $K_{Y_1} = K_{Y_0}$, the path-integral over $X^-$, that is (\ref{pi}), will, in this case, be given by 
\be
\Psi(\Phi_{Y_0}) = \int_{\Phi_{Y_0}} \CD \Phi \  e^{- {S_E}} \ \Psi(\Phi_{{\overline Y}_0}) =  \int_{\Phi_{Y_0}} \CD \Phi \  e^{- {S_E}}
\label{pi co}
\ee
(since as mentioned above, specifying the boundary values of the fields on ${\overline Y}_0$ is equivalent to specifying those on its mirror ${Y}_0$). Because (\ref{pi co}) is a path-integral without operator insertions that, as explained earlier, is also invariant under metric deformations of $X$, we can compute it as  $e^{- Ht}$~\cite{MS} in the limit $t \to \infty$,  where $t$ is the (stretched) interval of $X^-$. However, since $H$ acts as zero on the $\cq$-cohomology classes, (\ref{pi co}) is always equal to $1$ in our context; in other words, if we label $\cal H$ in (\ref{homo H}) as ${\cal H}(Y_1; K_{Y_1},  Y_0; K_{Y_0})$, we have from (\ref{pi co})
\be
{\cal H}(Y_0; K_{Y_0},  Y_0; K_{Y_0}) = 1;
\label{id}
\ee
a relation which can be thought to arise from a trivial cobordism from $({Y}_0, K_{{Y}_0})$ to $(Y_0, K_{Y_0})$ furnished by $(X^-, \Sigma^-)$. This result is also part of Proposition 3.27 in~\cite{dym}.

\bigskip\noindent{\it Composition Of Cobordisms And Maps} 

Let us consider the setup in fig.~3 again, but now, with the boundary values of the fields on $Y_1$ \emph{determined}. What we would like to do next is to compute the path integral over the segments spanned by $X^-$ and $X'$. Note that from the general ideas of quantum field theory, one can either compute this as a single path-integral starting from ${\overline Y}_0$ and ending at $Y_2$, or as a path-integral over $X^-$ --  starting at ${\overline Y}_0$ and ending at $Y_1$ -- followed by another path-integral over $X'$ --  starting at ${\overline Y}_1$ and ending at $Y_2$. Consequently, we can write
\be
  \int_{\Phi_{Y_2}} \CD \Phi \  e^{- {S_E}} \ \Psi(\Phi_{{\overline Y}_1}) \cdot  \int_{\Phi_{Y_1}} \CD \Phi \  e^{- {S_E}} \ \Psi(\Phi_{{\overline Y}_0}) = \int_{\Phi_{{Y}_2}} \CD \Phi \  e^{- {S_E}} \ \Psi(\Phi_{{\overline Y}_0}).
\label{homo rewrite}
\ee 
 By the fact that the Hilbert spaces ${\mathscr H}_{\overline B}$ and ${\mathscr H}_B$ are canonically dual to each other whence one can identify $\Psi(\Phi_{\overline B})$ with $\Psi(\Phi_B)$, (\ref{homo rewrite}) will then mean that
\be
{\cal H}(Y_2; K_{Y_2},  Y_1; K_{Y_1}) \cdot {\cal H}(Y_1; K_{Y_1},  Y_0; K_{Y_0}) = {\cal H}(Y_2; K_{Y_2},  Y_0; K_{Y_0});
\label{cobord}
\ee
a relation which can be thought to arise from a composite cobordism from  $({Y}_0, K_{{Y}_0})$ to $(Y_1, K_{Y_1})$ to  $({Y}_2, K_{{Y}_2})$ furnished by $(X^-, \Sigma^-)$ and $(X', \Sigma')$, respectively. This result is also part of Proposition 3.27 in~\cite{dym}.  

If we let $(Y_2, K_{Y_2})  = (Y_0, K_{Y_0})$, then (\ref{id}) and (\ref{cobord}) will imply that
\be
 {\cal H}(Y_1; K_{Y_1},  Y_0; K_{Y_0}) = {\cal H}^{-1}(Y_0; K_{Y_0},  Y_1; K_{Y_1}),
\ee 
i.e., $\cal H$ is invertible and therefore, it is also an\emph{ isomorphism}.

\newsubsection{A Vanishing Theorem For  The Monopole Floer Homology Of Three-Manifolds}

Consider $X = M \times {\bf S}^1$ to be \emph{symplectic} with $b^+_2(X) = b_1(M) > 1$, where $M$ is a compact, oriented three-manifold. Recall from our discussions in $\S$4 that in our case, ${\rm Gr}(c_1({\cal E}))$ only counts (with signs) pseudo-holomorphic curves $\Sigma$ -- with Poincar\'e-dual  $c_1(\cal E)$ -- which are nontrivially-embedded in $X$ such that $c_1({\cal E})\cdot [\omega_{\rm sp}] > 0$, where $[\omega_{\rm sp}] $ is the Poincar\'e-dual of the symplectic two-form $\omega_{\rm sp}$ on $X$. Consequently,  ${\rm Gr}(c_1({\cal E})) = 0$ \emph{identically} if $c_1({\cal E})\cdot [\omega_{\rm sp}] \leq 0$, and by (\ref{wow}) and (\ref{chi=chi}), it will mean that the monopole Floer homology groups $HM_\ast(M, \pi (\hat {\frak s}))$ ought to vanish if $c_1({\cal E})\cdot [\omega_{\rm sp}] \leq 0$;  here, $\pi: M \times {\bf S}^1 \to M$, and the first Chern class of the $\spin$-structure $\hat {\frak s}$ on $X$ is given by $2c_1(\hat {\frak s}) = 2 c_1({\cal E}) - c_1(K)$, where $K$ is the canonical line bundle on $X$. Note that this easy-to-reach but nevertheless important conclusion about $HM_\ast(M, \pi (\hat {\frak s}))$ has also been derived via a distinct and highly-involved mathematical approach in the Main Theorem of~\cite{SWF}, where ``$e$'' and ``${\frak s}_e$'' therein correspond to $\pi(c_1({\cal E}))$ and $\pi(\hat {\frak s})$ herein.

\bigskip\noindent{\it Mathematical Versus Physical Computation}

In the mathematical proof of the Main Theorem in~\cite{SWF} by Kutluhan and Taubes, the above conclusion about the vanishing of $HM_\ast(M, \pi (\hat {\frak s}))$ was obtained via a head-on analysis of the three-dimensional SW equations on $M$. In particular, the equations were checked for the presence or absence of sensible solutions (which directly generate  $HM_\ast(M, \pi (\hat {\frak s}))$)   under various conditions; no reference to other related invariants of $M$ or $X$ were made at all.   

On the other hand, in our computation leading to the above conclusion, we relied solely on the physically-derived relations (\ref{wow}) and (\ref{chi=chi}) -- which connect the topological invariants in various dimensions to one another -- without appealing to the three-dimensional SW equations on $M$. Thus, our physical computation provides, in this manner, a completely new way of deriving and understanding the vanishing of  $HM_\ast(M, \pi (\hat {\frak s}))$ when $c_1({\cal E})\cdot [\omega_{\rm sp}] \leq 0$.

\newsubsection{Seiberg-Witten Invariants Determined By The Canonical Basic Class}

Let $X$ be  a compact, oriented, symplectic four-manifold with $b_1 =0$ and $b^+_2 > 1$. Recall from (\ref{SW = SW}) and (\ref{exp SW}) that for $\frak s = \half c_1(K)$, we have
\be
SW(\frak s - \delta_{\cal C}) = SW (\frak s)
\label{SWs}
\ee 
if and only if $d_K = -c_1(K)[{\cal C}] + {\cal C} \cap {\cal C} = 0$, and according to (\ref{d condition}),  $\cC$ can be a pseudo-holomorphic curve in $X$. 

As explained earlier, (\ref{GT 3}) implies that $c_1(K)$ is the Poincar\'e dual of some pseudo-holomorphic curve in $X$. Let ${\cal C}$ be such a curve, i.e., $\delta_{\cal C} = {1\over 2} c_1(K^2)$;  as required, $d_K = 0$. Substituting this in (\ref{SWs}), we find that
\be
SW(\frak s_c) = SW(\frak s),
\label{sws}
\ee 
where $\frak s_c = {1\over 2}c_1(K^{-1})$ is the canonical $\spin$-structure.

Hence, from (\ref{GT 1}), (\ref{Gr =1}) and (\ref{sws}), we conclude that
\be
SW(\frak s_c) = +1
\label{sw=1}
\ee
on $X$. This easy-to-reach but nevertheless important conclusion about $SW(\frak s_c)$ has also been proved via a highly-involved and distinct mathematical approach in Proposition 2.1 of article 4 in~\cite{CT 1}. 

\bigskip\noindent{\it Mathematical Versus Physical Computation}

In the mathematical computation of (\ref{sw=1}) in Proposition 2.1 of article 4 in~\cite{CT 1},  the magnitude of $SW(\frak s_c)$ is determined to be unity because the SW equations are shown to have a unique solution; the positive sign arises because the kernel of an elliptic operator associated with the linearization of the equations, is trivial.  

On the other hand, our physical computation of (\ref{sw=1}) depends on the following: firstly, (\ref{sws}), which leads to (\ref{sw=1}), is a consequence of  (\ref{GT 3}) which implies that there is a pseudo-holomorphic curve in $X$ that is Poincar'e dual to $c_1(K)$; secondly, the positive sign in (\ref{sw=1}) can be seen to originate from  (\ref{Gr =1}), i.e., the fact that the number of connected, non-multiply-covered, pseudo-holomorphic curves $\Sigma \subset X$ of positive self-intersection which are ``positively-oriented'' is greater than the number which are ``negatively-oriented''  by one. Thus, our physical computation provides, in this manner, a completely new way of deriving and understanding  (\ref{sw=1}).

\newsubsection{About The Seiberg-Witten Invariants Of K\"ahler Manifolds}

What if $X$ in $\S$7.3 is K\"ahler? Then, one can say the following.  First, since every compact, oriented, K\"ahler manifold is necessarily symplectic, (\ref{sw=1}) will apply to $X$ as well. This observation is just Theorem 3.3.2 of~\cite{nico}.  Second, on any K\"ahler manifold such as $X$ where the almost complex structure $J$ is necessarily integrable, all points in the space $H$ of pseudo-holomorphic curves have positive orientation; i.e., all points in $H$ contribute as $+1$ in the computation of ${\rm Gr}(c_1(\cal E))$~\cite{dusa}.  In light of (\ref{Gr =1}) -- i.e., ${\rm Gr}(c_1({\cal E})) = + 1$  -- this means that $H$ consists of a\emph{ single} point only.  Third, note that via (\ref{sws}) and (\ref{GT 1}), we have $SW({\frak s}_c) =  {\rm Gr}(c_1({\cal E}))$. Therefore,  this means that from each solution of the ordinary SW equations on $X$ determined by ${\frak s}_c$, one can derive a pseudo-holomorphic curve in $X$ whose fundamental class is Poincar\'e dual to $c_1(\cal E)$. In other words,  the number of points in the moduli space $\CM_{{\frak s}_c}$ of solutions of the ordinary SW equations determined by ${\frak s}_c$  equals  the number of points in $H$. Thus, according to the second statement above, $\CM_{{\frak s}_c}$ consists of  a \emph{single} point only. This easy-to-reach but nevertheless important conclusion about $\CM_{{\frak s}_c}$ has also been proved via a highly-involved and distinct mathematical approach in Proposition 3.3.1 of~\cite{nico}.  

\bigskip\noindent{\it Mathematical Versus Physical Computation}

In the mathematical computation of (\ref{sw=1}) for K\"ahler manifolds in Theorem 3.3.2 of~\cite{nico},  the magnitude of $SW(\frak s_c)$ is again determined to be unity because the SW equations are shown in Proposition 3.3.1 of~\cite{nico} to have a unique solution; the positive sign arises because a relevant map between vector spaces defined by a  certain  ``resonance operator'' is orientation-preserving.  

On the other hand, the basis of our physical computation of (\ref{sw=1}) for symplectic and thus K\"ahler manifolds, is as described at the end of the previous subsection. Moreover, in the case where $X$ is K\"ahler, one can, from our physical computation, understand the uniqueness of the solution of the SW equations to be a consequence of the fact that there is just \emph{one}, connected, non-multiply-covered pseudo-holomorphic curve of positive self-intersection in $X$ that is nontrivial in homology. Thus, our physical computation provides, in this manner, a completely new way of deriving and understanding the SW invariants of ``admissible'' K\"ahler four-manifolds.

\newpage

\vspace{0.0cm}
\hspace{-1.0cm}{\large \bf Acknowledgements:}\\ 
I would like to thank A.J. Berrick, O. Collin, F. Han and Y.L. Wong for mathematical consultations. This work is supported in part by a start-up grant from the National University of Singapore. 
\vspace{0.0cm}



\end{document}